\pdfoutput=1
\documentclass{sn-jnl}
\usepackage[T1]{fontenc}
\usepackage[utf8]{inputenc}

\usepackage{enumitem}

\usepackage{bussproofs}
\usepackage{fancyvrb}
\usepackage{relsize}
\usepackage{cleveref}
\usepackage{listing}
\newcommand{\srclink}[1]{\hfill\href{(sources)}{https://archive.softwareheritage.org/#1}}
\newcommand{\srclinkfoot}[1]{\href{(sources)}{https://archive.softwareheritage.org/#1}}

\usepackage[all,2cell,cmtip]{xy}
\UseAllTwocells

\usepackage{hyperref}
\hypersetup{colorlinks=true}

\jyear{2021}%

\theoremstyle{thmstyleone}%
\newtheorem{theorem}{Theorem}
\theoremstyle{thmstyletwo}%

\theoremstyle{thmstylethree}%

\normalbaroutside  

\usepackage{academicons}
\definecolor{orcidlogocol}{HTML}{A6CE39}

\begin{document}
\title{Mechanising G\"odel-L\"ob provability logic in HOL Light}

\author[1]{\fnm{Marco} \sur{Maggesi}}\email{marco.maggesi@unifi.it}\equalcont{Affiliated with INdAM-GNSAGA}%
\affil[1]{\orgdiv{} \orgname{University of Florence}, \orgaddress{\state{Tuscany}, \country{Italy}}%
}

\author[2]{\fnm{Cosimo} \sur{Perini} \sur{Brogi}}\email{cosimo.perinibrogi@imtlucca.it}\equalcont{Affiliated with INdAM-GNSAGA}%
\affil[2]{\orgname{} \orgname{IMT School for Advanced Studies Lucca}, \orgaddress{\state{Tuscany}, \country{Italy}}%
}



\abstract{We introduce our implementation in HOL Light of the metatheory for G\"odel-L\"ob provability logic (GL), covering soundness and completeness w.r.t.\ possible world semantics and featuring a prototype of a theorem prover for GL itself. 
The strategy we develop here to formalise the modal completeness proof overcomes the technical difficulty due to the non-compactness of GL and is an adaptation -- according to the formal language and tools at hand -- of the proof given in George Boolos' 1995 monograph.
Our theorem prover for GL relies then on this formalisation, is implemented as a tactic of HOL Light that mimics the proof search in the labelled sequent calculus $\mathsf{G3KGL}$, and works as a decision algorithm for the provability logic: if the algorithm positively terminates, the tactic succeeds in producing a HOL Light theorem stating that the input formula is a theorem of GL; if the algorithm negatively terminates, the tactic extracts a model falsifying the input formula.

We discuss our code for the formal proof of modal completeness and the design of our proof search algorithm. Furthermore, we propose some examples of the latter's interactive and automated use.\footnote{This version of the article has been superseded by the Version of Record published in the \emph{Journal of Automated Reasoning} and does not reflect post-acceptance improvements, or any corrections. The Version of Record is available online at: \url{https://doi.org/10.1007/s10817-023-09677-z}.}}


\maketitle

\section{Introduction}
\VerbatimFootnotes

The origin of provability logic dates back to a short paper by G\"odel \cite{godel1933} where propositions about provability are formalised through a unary operator $\mathtt{B}$ to give a classical reading of intuitionistic logic.
That work opened the question of finding an adequate modal calculus for the formal properties of the provability predicate used in G\"odel's incompleteness theorems.
The problem has been settled since the 1970s for many formal systems of arithmetic employing G\"odel-L\"ob logic GL: the abstract properties of the formal provability predicate of any $\Sigma_{1}$-sound arithmetical theory $T$ extending $\mathsf{I\Sigma_{1}}$ \cite{sep-logic-provability} are captured by that modal system, as established by Solovay \cite{solovay1976provability}.

Solovay's technique consists of an arithmetization of a relational countermodel for a given formula that is not a theorem of $\mathbb{GL}$, from which it is possible to define an appropriate arithmetical formula that is not a theorem of the mathematical system.\footnote{
In contemporary research, this is still the main strategy to prove arithmetical completeness for other modalities for provability and related concepts, particularly for interpretability logics.}

Therefore, on the one hand, completeness of formal systems w.r.t.~the relevant relational semantics is still an unavoidable step in achieving the more substantial result of arithmetical completeness; on the other hand, however, the area of provability logic keeps flourishing and suggesting old and new open problems.\footnote{Some of them are closely related to the field of proof theory; others point at developing a uniform proof-strategy to establish adequate semantics in formal theories of arithmetic having different strengths and flavours. The reader is referred to Beklemishev and Visser \cite{beklemishev-visser} for a survey of open problems in provability logics. 
}

\bigskip

Our work starts then with a deep embedding of the syntax of propositional modal logic together with the corresponding relational semantics.
Next, we introduce the traditional axiomatic calculus $\mathbb{GL}$ and prove the soundness of the system w.r.t.~irreflexive transitive finite frames.

A more mathematical part then follows: 
our goal has been proving formally

\begin{theorem}\label{completezza}
For any formula $A$, $\mathbb{GL}\vdash A$ iff $A$ is true in any irreflexive transitive finite frame.
\end{theorem}

Since GL is not compact, the standard methodology based on canonical models \cite{popkorn1994first} cannot be directly applied here. After Kozen and Parikh \cite{KOZEN1981113}, it is common to restrict the construction to a finite subformula universe.\footnote{We are grateful to an anonymous referee for pointing us to that work.} The same idea is basically used in Boolos \cite{boolos1995logic} to prove modal completeness for GL.
Proceeding in that same way, we have to formally verify a series of preliminary lemmas and constructions involving the behaviour of syntactical objects used in the standard proof of the completeness theorem. These unavoidable steps are often only proof-sketched in wide-adopted textbooks in logic -- including \cite{boolos1995logic} -- for they mainly involve ``standard'' reasoning \emph{within} the proof system we are dealing with. Nevertheless, when working in a formal setting, as we did with HOL Light, we need to split down the main goal into several subgoals, dealing with both the object- and the meta-level. Sometimes, the HOL Light automation does mirror -- and, using automation mechanisms, simplify details of --  the informal reasoning. On other occasions, we have to modify some aspects of the proof strategies, simplifying, through the computer tools at hand, some passages of the informal argument in Boolos \cite[Ch.~5]{boolos1995logic}.

\bigskip

As it is known, for any logical calculus, a completeness result w.r.t.~finite models --aka finite model property-- implies the decidability of that very logic \cite{harrop_1958}. Therefore, the formal proof for Theorem \ref{completezza} we discuss in the first part of the present work could be used, in principle, to develop a decision algorithm for $\mathbb{GL}$.
That would be a valuable tool for automating in HOL Light the proof of theorems of GL.

Proof search in axiomatic calculi is a challenging task. In our formalisation, we had to develop several proofs in the axiomatic calculus for $\mathbb{GL}$, and only in a few cases, it has been possible to leave the proof search to the automation mechanism of HOL Light.

By having a formal proof of the finite model property, one could hope to solve the goal of checking whether a formula is a theorem of GL by shifting from the syntactic problem of finding a proof of that formula in the axiomatic calculus to the semantic problem of checking the validity of that formula in any finite model by applying automated first-order reasoning as implemented in the proof assistant.

Such a strategy has many shortcomings, unfortunately.
We document some of its limits in Section \ref{sec:bisim} and at the beginning of the subsequent section, but the main points are the following: From the complexity theory viewpoint, a decision procedure based on the countermodel construction would be far from being optimal, belonging to \textsc{EXPSPACE}, rather than \textsc{PSPACE}, to which decidability of GL belongs \cite{ladner1977computational};\footnote{Refer to \ref{sec:bisim} below for the explicit bound one can obtain for the complexity of such a decision procedure.} from the practical viewpoint, implementing it in HOL Light would consists in a relatively simple formalisation exercise.

Structural proof theory for modal logics -- briefly summarised in Section \ref{sec:bits} -- suggests a more promising strategy.

\medskip

Many contemporary sequent calculi for non-classical logics are based on an ``internalisation'' of possible world semantics in Gentzen's original formalism \cite{negri2011proof,poggiolesi2010gentzen}.

We resort to an explicit internalisation for developing our decision algorithm since our formalisation in HOL Light of Kripke semantics for GL is \emph{per se} a labelling technique in disguise.

Therefore, in the second part of this paper, we introduce what might be considered a shallow embedding in HOL Light of Negri's labelled sequent calculus $\mathsf{G3KGL}$ \cite{negri2005proof,negri2014proofs}.

To state it clearly: while in the first part, we defined the axiomatic system $\mathbb{GL}$ within our proof assistant employing an inductive definition of the derivability relation for $\mathbb{GL}$, in the second part, we define \emph{new tactics} of HOL Light in order to perform a proof search in $\mathsf{G3KGL}$ by using the automation infrastructure provided by HOL Light itself.

By relying on the meta-theory for $\mathsf{G3KGL}$ developed in \cite{negri2014proofs}, we can safely claim that such an embedding provides a decision algorithm for GL: if proof search terminates on a modal formula given as input, HOL Light produces a new theorem, stating that the input formula is a theorem of GL; otherwise, our algorithm prints all the information necessary to construct an appropriate countermodel for the input formula.

Our code is integrated into the current HOL Light distribution and freely available from \cite{hol-light}. The files we will occasionally refer to during the presentation are freely accessible from there.

\bigskip



The present paper is structured as follows:
\begin{itemize}
\item In Section \ref{synsem}, we introduce the basic ingredients of our development, namely the formal counterparts of the syntax and relational semantics for provability logic, along with some lemmas and general definitions which are useful to handle the implementation of these objects uniformly, i.e.~without the restriction to the specific modal system we are interested in. The formalisation constitutes a large part of the file \verb|modal.ml|;
\item In Section \ref{axiom}, we formally define the axiomatic calculus $\mathbb{GL}$ and prove neatly the validity lemma for this system. Moreover, we give formal proofs of several lemmas \emph{in} $\mathbb{GL}$ ($\mathbb{GL}$-lemmas, for short), whose majority is, in fact, common to all normal modal logics, so that our proofs might be re-used in subsequent implementations of different systems. This corresponds to the contents of our code in \verb|gl.ml|;
\item In Section \ref{compl} we give our formal proof of modal completeness of $\mathbb{GL}$, starting with the definition of maximal consistent \emph{lists} of formulas. In order to prove their syntactic properties -- and, in particular, the extension lemma for consistent lists of formulas to maximal consistent lists -- we use the $\mathbb{GL}$-lemmas 
and at the same time, we adapt an already known general proof-strategy to maximise the gain from the formal tools provided by HOL Light -- or, informally, from higher-order reasoning.

At the end of the section, we give the formal definition of bisimilarity for our setup, and we prove the associated bisimulation theorem \cite[Ch.~11]{popkorn1994first}. Our notion of bisimilarity is polymorphic because it can relate classes of frames sitting on different types.
With this tool at hand, we can correctly state our completeness theorem in its natural generality (\verb|COMPLETENESS_THEOREM_GEN|) -- i.e.~for irreflexive, transitive finite frames over any (infinite) type -- this way obtaining the finite model property for GL w.r.t.~frames having finite \emph{sets} of formulas as possible worlds, as in the standard Henkin's construction.
These results conclude the first part of the paper and are gathered in the file \verb|completeness.ml|.

\item Section \ref{prooftheorygl} opens the part of the paper describing our original theorem prover for GL. We collect some basic notions and techniques for proof-theoretic investigations on modal logic. In particular, we recall  the methodology of explicit internalisation of relational semantics and the results that provide the meta-theory for our decision algorithm.

\item Finally, in Section \ref{decision}, we describe our implementation of the labelled sequent calculus $\mathsf{G3KGL}$ -- documented by the file \verb|decid.ml| -- to give a decision procedure for GL. We recover the formalisation of Kripke semantics for GL presented in Section \ref{synsem} to define new tactics mimicking the rules for $\mathsf{G3KGL}$. Then, we properly define our decision algorithm by designing a specific terminating proof search strategy in the labelled sequent calculus. This way, we extend the HOL Light automation toolbox with an ``internal'' theorem prover for GL that can also produce a countermodel for any formula for which proof search fails. We propose some hands-on examples of use by considering modal principles that have a certain relevance for meta-mathematical investigations; the interested reader will find further examples in the file \verb|tests.ml|.

\item Section \ref{relwork} closes the paper and compares our results with related works on mechanised modal logic.

\end{itemize}

\smallskip

Our formalisation does not modify any original HOL Light tools, and it is therefore ``foundationally safe''. Moreover, since we only used that original formal infrastructure, our results can be easily translated into another theorem prover belonging to the HOL family endowed with the same automation toolbox.



\paragraph*{Revision notes.} This article is an expanded version of the conference paper \cite{maggesi_et_al:LIPIcs.ITP.2021.26},
presented at Interactive Theorem Proving (ITP) 2021. Changes include adding Sections \ref{prooftheorygl} and \ref{decision} on our implementation in HOL Light of the theorem prover and countermodel constructor for GL, and some minor local revisions. An intermediate version of the contents discussed in the present paper appeared in \cite{11567_1091313}.
Both authors wish to thank two anonymous referees for their valuable feedback and comments towards improving the presentation of the material we discuss here.

\subsection{HOL Light notation}
\label{sec:hol-light-notation}

The HOL Light proof assistant \cite{hol-light} is based on \emph{classical} higher-order logic with polymorphic type variables and where equality is the only primitive notion. From a logical viewpoint, the formal engine defined by the \emph{term-conversions} and \emph{inference rules} underlying HOL Light is the same as that described in \cite{lambekscott}, extended by an infinity axiom and the classical characterization of Hilbert's choice operator. From a practical perspective, it is a theorem prover privileging a procedural proof style development. I.e., when using it, we have to solve goals by applying \emph{tactics} that reduce them to (hopefully) simpler subgoals so that the interactive aspect of proving is highlighted. Proof scripts can then be constructed using \emph{tacticals} that compact the proof into a few lines of code evaluated by the machine.

In what follows, we will report several code fragments to give the flavour of our development and to provide additional documentation and information to the reader interested in the technical details.
We partially edited the code to ease the reading of the mathematical formulas.
For instance, we replaced the purely ASCII notations of HOL Light with the usual graphical notation.
Every source code snippet in this paper has a link ---indicated by the word ``(sources)'' on the right---  to a copy of the source files stored on the Software Heritage\footnote{\href{https://www.softwareheritage.org/}{www.softwareheritage.org}} archive which is helpful to those who want to see the raw code and how it fits in its original context. Here is an example:
\begin{lstlisting}[mathescape=true,escapeinside={(*!}{!*)}]
  ADD_SYM(*! {\srclink{swh:1:cnt:e47e19813eec4cbd6ffd9d8082c86b0b421c9c2e;origin=https://github.com/jrh13/hol-light;visit=swh:1:snp:fe1b3b83e1dd2bde44d3698f723b25c949de2851;anchor=swh:1:rev:ab57c07ff0105fef75a9fcdd179eda0d26854ba3;path=/arith.ml;lines=67-69}} !*)
    $\vdash$ $\forall$m n. m + n = n + m
\end{lstlisting}
Note that we report theorems with their associated name (the name of its associated OCaml constant), and we write their statement prefixed with the turnstile symbol~($\vdash$).
In the expository style, we omit formal proofs, but the meaning of definitions, lemmas, and theorems in natural language is clear.

The HOL Light printing mechanism omits type information completely.  Therefore, we warn the reader about types when it might be helpful, or even indispensable, to avoid ambiguity -- including the case of our main results, \verb|COMPLETENESS_THEOREM| and \verb|COMPLETENESS_THEOREM_GEN|.

We also recall that a Boolean function \texttt{s~:\ $\alpha$~->~bool} is also called a \emph{set on $\alpha$} in the HOL parlance.  The notation \texttt{x~IN~s} is equivalent to \texttt{s~x} and must not be confused with a type annotation \texttt{x~:~$\alpha$}.

\bigskip

As mentioned, our contribution is part of the HOL Light distribution.
The reader interested in performing these results on her machine -- and perhaps building further formalisation on top of it -- can run our code with the command
\begin{lstlisting}
  loadt "GL/make.ml";;
\end{lstlisting}
at the HOL Light prompt.

\section{Basics of Modal Logic}\label{synsem}

As we stated, we deal with a logic that extends classical propositional reasoning using a single modal operator intended to capture the abstract properties of the provability predicate for arithmetic.

To reason about and within this logic, we have to ``teach'' HOL Light -- our meta-language -- how to identify it, starting with its syntax -- the object-language -- and semantics -- the interpretation of this very object-language.

From a foundational perspective, we want to keep everything neat and clean; therefore, we will define \emph{both} the object-language and its interpretation with no relation to the HOL Light environment. In other terms: our formulas and operators are \emph{real} syntactic objects which we keep distinct from their semantic counterparts -- and from the logical operators of the theorem prover too.

\subsection{Language and semantics defined}
\label{sec:lang-sem-def}

Let us start by fixing the propositional modal language $\mathcal{L}$ we will use throughout the present work. We consider \emph{all} classical propositional operators -- conjunction, disjunction, implication, equivalence, negation, along with the 0-ary symbols $\top$ and $\bot$ -- and add a modal unary connective $\Box$. The starting point is, as usual, a denumerable infinite set of propositional atoms $a_{0},a_{1},\cdots$.
Accordingly, formulas of this language will have one of the following forms
\[
  \bot\;\,|\;\, \top\;\, |\;\, a\;\, |\;\, \neg A\;\, |\;\,
  A\wedge B\,\, |\;\, A \vee B\;\, |\;\,
  A\rightarrow B\;\, |\;\, A\leftrightarrow B\;\,|\;\, \Box A\;\,.
\]
The following code extends the HOL system with
an \textbf{inductive type of formulas} generated by the above syntactic constructions
\begin{lstlisting}[escapeinside={(*!}{!*)}]
  let form_INDUCT,form_RECURSION = define_type(*! \srclink{swh:1:cnt:e96b1bf9513c5683d53e971912903fc13d21b282;origin=https://github.com/jrh13/hol-light;visit=swh:1:snp:fe1b3b83e1dd2bde44d3698f723b25c949de2851;anchor=swh:1:rev:ab57c07ff0105fef75a9fcdd179eda0d26854ba3;path=/GL/modal.ml;lines=21-30} !*)
    "form = False
          (*!\textbar!*) True
          (*!\textbar!*) Atom string
          (*!\textbar!*) Not form
          (*!\textbar!*) && form form
          (*!\textbar!*) (*!\textbar\textbar!*) form form
          (*!\textbar!*) --> form form
          (*!\textbar!*) <-> form form
          (*!\textbar!*) Box form";;
\end{lstlisting}
Later in the code, the operators \verb$&&$, \verb$||$, \verb$-->$, \verb$<->$ are used infixed as usual.
Next, we turn to semantics for our modal language. We use \textbf{relational models} -- aka Kripke models.\footnote{See Copeland \cite{copeland2002genesis} for the historical development of this notion.}

Formally, a \textbf{Kripke frame} is made of a non-empty set `of possible worlds' \verb|W|, together with a binary relation \verb$R$ on \verb$W$. To this, we add an evaluation function \verb$V$, which assigns to each atom of our language and each world \verb$w$ in \verb$W$ a Boolean value.
This is extended to a \textbf{forcing relation} \texttt{holds}, defined recursively on the structure of the input formula \verb$p$, that computes the truth-value of \verb$p$ in a specific world \verb$w$:
\begin{lstlisting}[mathescape=true,escapeinside={(*!}{!*)}]
  let holds = new_recursive_definition form_RECURSION(*! \srclink{swh:1:cnt:e96b1bf9513c5683d53e971912903fc13d21b282;origin=https://github.com/jrh13/hol-light;visit=swh:1:snp:fe1b3b83e1dd2bde44d3698f723b25c949de2851;anchor=swh:1:rev:ab57c07ff0105fef75a9fcdd179eda0d26854ba3;path=/GL/modal.ml;lines=40-55} !*)
    `(holds f V False (w:W) $\Longleftrightarrow$ $\bot$) $\wedge$
     (holds f V True w $\Longleftrightarrow$ $\top$) $\wedge$
     (holds f V (Atom s) w $\Longleftrightarrow$ V s w) $\wedge$
     (holds f V (Not p) w $\Longleftrightarrow$ $\neg$(holds f V p w)) $\wedge$
     (holds f V (p && q) w $\Longleftrightarrow$
      holds f V p w $\wedge$ holds f V q w) $\wedge$
     (holds f V (p (*!\textbar\textbar!*) q) w $\Longleftrightarrow$
      holds f V p w $\vee$ holds f V q w) $\wedge$
     (holds f V (p --> q) w $\Longleftrightarrow$
      holds f V p w $\Longrightarrow$ holds f V q w) $\wedge$
     (holds f V (p <-> q) w $\Longleftrightarrow$
      holds f V p w = holds f V q w) $\wedge$
     (holds f V ($\Box$ p) w $\Longleftrightarrow$
      $\forall$u. u $\in$ FST f $\wedge$ SND f w u $\Longrightarrow$ holds f V p u)`;;
\end{lstlisting}
In the previous lines of code, \verb|f| stands for a generic Kripke frame -- i.e.\ a pair \verb|(W:W->bool,R:W->W->bool)| of a set of worlds and an accessibility relation -- and \verb|V:string->W->bool| is an evaluation of propositional variables.
Then, the \textbf{validity} of a formula \verb|p| with respect to a frame \verb|(W,R)|, and a class of frames \verb|L|, denoted respectively \verb|holds_in (W,R) p| and \verb$L |= p$, are
\begin{lstlisting}[mathescape=true,escapeinside={(*!}{!*)}]
  let holds_in = new_definition(*! \srclink{swh:1:cnt:e96b1bf9513c5683d53e971912903fc13d21b282;origin=https://github.com/jrh13/hol-light;visit=swh:1:snp:fe1b3b83e1dd2bde44d3698f723b25c949de2851;anchor=swh:1:rev:ab57c07ff0105fef75a9fcdd179eda0d26854ba3;path=/GL/modal.ml;lines=57-58} !*)
    `holds_in (W,R) p $\Longleftrightarrow$
     $\forall$V w. w $\in$ W $\Longrightarrow$ holds (W,R) V p w`;;
\end{lstlisting}
\begin{lstlisting}[mathescape=true,escapeinside={(*!}{!*)}]
  let valid = new_definition(*! \srclink{swh:1:cnt:e96b1bf9513c5683d53e971912903fc13d21b282;origin=https://github.com/jrh13/hol-light;visit=swh:1:snp:fe1b3b83e1dd2bde44d3698f723b25c949de2851;anchor=swh:1:rev:ab57c07ff0105fef75a9fcdd179eda0d26854ba3;path=/GL/modal.ml;lines=62-63} !*)
    `L (*!\textbar!*)= p $\Longleftrightarrow$ $\forall$f. L f $\Longrightarrow$ holds_in f p`;;
\end{lstlisting}
The above formalisation is essentially presented in Harrison's HOL Light Tutorial~\cite[§~20]{harrisontutorial}. Notice that the usual notion of Kripke frame requires that the set of possible worlds is non-empty: that condition could be imposed by adapting the \verb|valid| relation. We have preferred to stick to Harrison's original definitions in our code. However, in the next section, when we define the classes of frames, we are dealing with for GL, the requirement on \verb|W| is correctly integrated with the corresponding types.

\subsection{Frames for GL}
\label{sec:frames-GL}
For carrying out our formalisation, we are interested in the logic of the (non-empty) frames whose underlying relation $R$ is \textbf{transitive} and conversely well-founded -- aka \textbf{Noetherian} -- on the corresponding set of possible worlds; in other terms, we want to study the modal tautologies in models based on an accessibility relation $R$ on $W$ such that
\begin{itemize}
\item if $xRy$ and $yRz$, then $xRz$; and
\item for no $X\subseteq W$ there are infinite $R$-chains $x_{0}Rx_{1}Rx_{2}\cdots$.
\end{itemize}
In HOL Light, \verb|WF R| states that \verb|R| is a well-founded relation: then, we express the latter condition as \verb|WF(\x y. R y x)|.
Here we see a recurrent motif in logic: defining a system from the semantic perspective requires non-trivial tools from the foundational point of view, for, in order to express the second condition, a first-order language is not enough. However, that is not an issue here since our underlying system is natively higher order:\footnote{We warn the reader about a potentially misleading notation. In the following statement, two interrelated mathematical objects occur, both denoted by \verb|W| for convenience: one is the type \verb|W|, and the other is the set \verb|W| on the former -- in a sense explained in the introduction about the HOL Light notation.}

\begin{lstlisting}[mathescape=true,escapeinside={(*!}{!*)}]
  let TRANSNT = new_definition(*! \srclink{swh:1:cnt:e96b1bf9513c5683d53e971912903fc13d21b282;origin=https://github.com/jrh13/hol-light;visit=swh:1:snp:fe1b3b83e1dd2bde44d3698f723b25c949de2851;anchor=swh:1:rev:ab57c07ff0105fef75a9fcdd179eda0d26854ba3;path=/GL/modal.ml;lines=99-104} !*)
    `TRANSNT (W:W->bool,R:W->W->bool) $\Longleftrightarrow$
     $\neg$(W = {}) $\wedge$
     ($\forall$x y:W. R x y $\Longrightarrow$ x $\in$ W $\wedge$ y $\in$ W) $\wedge$
     ($\forall$x y z:W. x $\in$ W $\wedge$ y $\in$ W $\wedge$ z $\in$ W $\wedge$ R x y $\wedge$ R y z
                $\Longrightarrow$ R x z) $\wedge$
     WF($\lambda$x y. R y x)`;;
\end{lstlisting}



We can characterize this class of frames by using a \emph{propositional} language extended by a modal operator $\Box$ that satisfies the \emph{G\"odel-L\"ob axiom schema} ($\mathsf{GL}$) $:\Box(\Box A\rightarrow A)\rightarrow\Box A.$
Here is the formal version of our claim:


\begin{lstlisting}[mathescape=true,escapeinside={(*!}{!*)}]
  TRANSNT_EQ_LOB(*! \srclink{swh:1:cnt:e42639c66fe78808c04d8bedbb9ce84131562084;origin=https://github.com/jrh13/hol-light;visit=swh:1:snp:fe1b3b83e1dd2bde44d3698f723b25c949de2851;anchor=swh:1:rev:ab57c07ff0105fef75a9fcdd179eda0d26854ba3;path=/GL/gl.ml;lines=121-129} !*)
    $\vdash$ $\forall$W:W->bool R:W->W->bool.
        ($\forall$x y:W. R x y $\Longrightarrow$ x $\in$ W $\wedge$ y $\in$ W)
        $\Longrightarrow$ (($\forall$x y z. x $\in$ W $\wedge$ y $\in$ W $\wedge$ z $\in$ W $\wedge$ R x y $\wedge$ R y z
                      $\Longrightarrow$ R x z) $\wedge$
             WF ($\lambda$x y. R y x) $\Longleftrightarrow$
             ($\forall$p. holds_in (W,R) (Box(Box p --> p) --> Box p)))
\end{lstlisting}

The informal proof of the above result is standard and can be found in \cite[Theorem 10]{boolos1995logic} and in \cite[Theorem 5.7]{popkorn1994first}.
The computer implementation of the proof is made easy thanks to Harrison's tactic \verb|MODAL_SCHEMA_TAC| for semantic reasoning in modal logic, documented in \cite[§~20.3]{harrisontutorial}.

Using this preliminary result, we could say that the frame property of being transitive and Noetherian can be captured by G\"odel-L\"ob modal axiom without recurring to a higher-order language.

Nevertheless, that class of frames is not particularly informative from a logical point of view: a frame in \verb|TRANSNT| can be too huge to be used, e.g., for mechanically checking whether it does provide a countermodel for a formula of our logic. In fact, when aiming at a completeness theorem, one wants to consider models that are helpful for establishing further properties of the same logic. In the present case, the decidability of GL, which, as for any other normal modal logic, is a straightforward corollary of the \emph{finite model property}~\cite[Ch.~13]{popkorn1994first}.


The frames we want to investigate are then those whose $W$ is \textbf{finite}, and whose $R$ is both \textbf{irreflexive} and \textbf{transitive}:

\begin{lstlisting}[mathescape=true,escapeinside={(*!}{!*)}]
  let ITF = new_definition(*! \srclink{swh:1:cnt:e42639c66fe78808c04d8bedbb9ce84131562084;origin=https://github.com/jrh13/hol-light;visit=swh:1:snp:fe1b3b83e1dd2bde44d3698f723b25c949de2851;anchor=swh:1:rev:ab57c07ff0105fef75a9fcdd179eda0d26854ba3;path=/GL/gl.ml;lines=152-158} !*)
    `ITF (W:W->bool,R:W->W->bool) $\Longleftrightarrow$
      $\neg$(W = $\emptyset$) $\wedge$
     ($\forall$x y:W. R x y $\Longrightarrow$ x $\in$ W $\wedge$ y $\in$ W) $\wedge$
     FINITE W $\wedge$
     ($\forall$x. x $\in$ W $\Longrightarrow$ $\neg$ R x x) $\wedge$
     ($\forall$x y z. x $\in$ W $\wedge$ y $\in$ W $\wedge$ z $\in$ W $\wedge$ R x y $\wedge$ R y z
      $\Longrightarrow$ R x z)`;;
\end{lstlisting}
Now it is easy to see that \texttt{ITF} is a subclass of \texttt{TRANSNT}:
\begin{lstlisting}[mathescape=true,escapeinside={(*!}{!*)}]
  ITF_NT(*! \srclink{swh:1:cnt:e42639c66fe78808c04d8bedbb9ce84131562084;origin=https://github.com/jrh13/hol-light;visit=swh:1:snp:fe1b3b83e1dd2bde44d3698f723b25c949de2851;anchor=swh:1:rev:ab57c07ff0105fef75a9fcdd179eda0d26854ba3;path=/GL/gl.ml;lines=160-167} !*)
    $\vdash$ $\forall$W R:W->W->bool. ITF(W,R) $\Longrightarrow$ TRANSNT(W,R)
\end{lstlisting}
That will be the class of frames whose logic we are now going to define syntactically.

\section{Axiomatizing GL}\label{axiom}

We want to identify the logical system generating all and only the modal tautologies for transitive Noetherian frames; more precisely, we want to isolate the \emph{generators} of the modal tautologies in the subclass of transitive Noetherian frames which are finite, transitive, and irreflexive.\footnote{Notice that the lemma \verb|ITF_NT| allows us to derive the former result as a corollary of the latter.} 

When dealing with the very notion of tautology -- or \emph{theoremhood}, discarding the complexity or structural aspects of \emph{derivability} in a formal system -- it is convenient to focus on axiomatic calculi. The calculus we deal with here is usually denoted by $\mathbb{GL}$.

It is clear from the definition of the forcing relation that for classical operators, any axiomatization of propositional classical logic will do the job.
Here, we adopt a basic system in which only $\rightarrow$ and $\bot$ are primitive -- from the axiomatic perspective -- and all the remaining classical connectives are defined by axiom schemas and by the inference rule of Modus Ponens imposing their standard behaviour.

Therefore, to the classical engine, we add
\begin{itemize}
\item the axiom schema $\mathsf{K}$:\quad $\Box(A\rightarrow B)\rightarrow\Box A\rightarrow\Box B$;
\item the axiom schema $\mathsf{GL}$:\quad $\Box(\Box A\rightarrow A)\rightarrow\Box A$;
\item the necessitation rule $\mathsf{NR}$: \ \AxiomC{$A$}\RightLabel{$\mathsmaller{\mathsf{NR}}$}\UnaryInfC{$\Box A$}\DisplayProof ,
\end{itemize}
where $A,B$ are generic formulas (not simply atoms).
Then, here is the complete definition of the \textbf{axiom system} $\mathbb{GL}$.
The inductive predicate \verb|GLaxiom| encodes the set of axioms for GL:
\begin{lstlisting}[mathescape=true,escapeinside={(*!}{!*)}]
  let GLaxiom_RULES,GLaxiom_INDUCT,GLaxiom_CASES =(*! \srclink{swh:1:cnt:e42639c66fe78808c04d8bedbb9ce84131562084;origin=https://github.com/jrh13/hol-light;visit=swh:1:snp:fe1b3b83e1dd2bde44d3698f723b25c949de2851;anchor=swh:1:rev:ab57c07ff0105fef75a9fcdd179eda0d26854ba3;path=/GL/gl.ml;lines=11-23} !*)
    new_inductive_definition
    `($\forall$p q. GLaxiom (p --> (q --> p))) $\wedge$
     ($\forall$p q r.
        GLaxiom ((p --> q --> r) --> (p --> q) --> (p --> r))) $\wedge$
     ($\forall$p. GLaxiom (((p --> False) --> False) --> p)) $\wedge$
     ($\forall$p q. GLaxiom ((p <-> q) --> p --> q)) $\wedge$
     ($\forall$p q. GLaxiom ((p <-> q) --> q --> p)) $\wedge$
     ($\forall$p q. GLaxiom ((p --> q) --> (q --> p) --> (p <-> q))) $\wedge$
     GLaxiom (True <-> False --> False) $\wedge$
     ($\forall$p. GLaxiom (Not p <-> p --> False)) $\wedge$
     ($\forall$p q.
        GLaxiom (p && q <-> (p --> q --> False) --> False)) $\wedge$
     ($\forall$p q. GLaxiom (p || q <-> Not(Not p && Not q))) $\wedge$
     ($\forall$p q. GLaxiom (Box (p --> q) --> Box p --> Box q)) $\wedge$
     ($\forall$p. GLaxiom (Box (Box p --> p) --> Box p))`;;
\end{lstlisting}
The judgment $\mathbb{GL} \vdash A$, denoted \verb$|-- A$ in the machine code (not to be confused with the symbol for HOL theorems $\vdash$), is also inductively defined in the expected way:\footnote{Small modifications to limit the application of $\mathsf{NR}$ on the definition of \verb+|--+ would introduce the notion of derivability from a set of assumptions, so that the deduction theorem would hold \cite{hakli2012does}.}
\begin{lstlisting}[mathescape=true,escapeinside={(*!}{!*)}]
  let GLproves_RULES,GLproves_INDUCT,GLproves_CASES =(*! \srclink{swh:1:cnt:e42639c66fe78808c04d8bedbb9ce84131562084;origin=https://github.com/jrh13/hol-light;visit=swh:1:snp:fe1b3b83e1dd2bde44d3698f723b25c949de2851;anchor=swh:1:rev:ab57c07ff0105fef75a9fcdd179eda0d26854ba3;path=/GL/gl.ml;lines=29-32} !*)
    new_inductive_definition
    `($\forall$p. GLaxiom p $\Longrightarrow$ (*!\textbar!*)-- p) $\wedge$
     ($\forall$p q. (*!\textbar!*)-- (p --> q) $\wedge$ (*!\textbar!*)-- p $\Longrightarrow$ (*!\textbar!*)-- q) $\wedge$
     ($\forall$p. (*!\textbar!*)-- p $\Longrightarrow$ (*!\textbar!*)-- (Box p))`;;
\end{lstlisting}

\subsection{Soundness lemma}

We can now prove that $\mathbb{GL}$ is \textbf{sound} -- i.e.~every formula derivable in the calculus is a tautology in the class of irreflexive transitive finite frames.
This result is obtained by simply unfolding the relevant definitions and applying theorems \verb|TRANSNT_EQ_LOB| and \verb|ITF_NT| of Section~\ref{sec:frames-GL}:


\begin{lstlisting}[mathescape=true,escapeinside={(*!}{!*)}]
  GL_TRANSNT_VALID(*! \srclink{swh:1:cnt:e42639c66fe78808c04d8bedbb9ce84131562084;origin=https://github.com/jrh13/hol-light;visit=swh:1:snp:fe1b3b83e1dd2bde44d3698f723b25c949de2851;anchor=swh:1:rev:ab57c07ff0105fef75a9fcdd179eda0d26854ba3;path=/GL/gl.ml;lines=143-146} !*)
    $\vdash$ $\forall$p. ((*!\textbar!*)-- p) $\Longrightarrow$ TRANSNT (*!\textbar!*)= p
\end{lstlisting}

\begin{lstlisting}[mathescape=true,escapeinside={(*!}{!*)}]
  GL_ITF_VALID(*! \srclink{swh:1:cnt:e42639c66fe78808c04d8bedbb9ce84131562084;origin=https://github.com/jrh13/hol-light;visit=swh:1:snp:fe1b3b83e1dd2bde44d3698f723b25c949de2851;anchor=swh:1:rev:ab57c07ff0105fef75a9fcdd179eda0d26854ba3;path=/GL/gl.ml;lines=169-174} !*)
    $\vdash$ $\forall$p. (*!\textbar!*)-- p $\Longrightarrow$ ITF (*!\textbar!*)= p
\end{lstlisting}
From this, we get a model-theoretic proof of \textbf{consistency for the calculus}
\begin{lstlisting}[mathescape=true,escapeinside={(*!}{!*)}]
  GL_consistent(*! \srclink{swh:1:cnt:e42639c66fe78808c04d8bedbb9ce84131562084;origin=https://github.com/jrh13/hol-light;visit=swh:1:snp:fe1b3b83e1dd2bde44d3698f723b25c949de2851;anchor=swh:1:rev:ab57c07ff0105fef75a9fcdd179eda0d26854ba3;path=/GL/gl.ml;lines=176-182} !*)
    $\vdash$ $\neg$((*!\textbar!*)-- False)
\end{lstlisting}
We are now ready to consider the mechanised proof of completeness for the calculus w.r.t.~this very class of frames. 

\subsection{\texorpdfstring{$\mathbb{GL}$}{GL}-lemmas}
\label{sec:GL-lemmas}
Proving some lemmas in the axiomatic calculus $\mathbb{GL}$ is a technical interlude necessary for obtaining the completeness result.

Following this aim, we denoted the classical axioms and rules of the system as the propositional schemas used by Harrison in the file \verb|Arithmetic/derived.ml| of the HOL Light standard distribution~\cite{hol-light} -- where, in fact, many of our lemmas relying only on the propositional calculus are already proven there w.r.t.~an axiomatic system for first-order classical logic; our further lemmas involving modal reasoning are denoted by names that are commonly used in informal presentations.

Therefore, the code in \texttt{gl.ml} mainly consists of the formalised proofs of those lemmas \emph{in} $\mathbb{GL}$ that are useful for the formalised results we present in the next section.
This file might be considered a ``kernel'' for further experiments in reasoning about axiomatic calculi using HOL Light. The lemmas we proved are, indeed, standard tautologies of classical propositional logic, along with specific theorems of minimal modal logic and its extension for transitive frames -- i.e.~of the systems $\mathbb{K}$ and $\mathbb{K}4$ \cite{popkorn1994first} --, so that by applying minor changes in basic definitions, they are -- so to speak -- take-away proof scripts for extensions of that very minimal system within the realm of normal modal logics.

Whenever it was useful, we have also given a characterisation of classical operators in terms of an implicit (i.e.~internal) deduction expressed by the connective \verb+-->+.  When this internal deduction is from an empty set of assumptions, we named the HOL theorem with the suffix \verb|_th|, and stated the deduction as a $\mathbb{GL}$-lemma, such as
\begin{lstlisting}[mathescape=true,escapeinside={(*!}{!*)}]
  GL_modusponens_th(*! \srclink{swh:1:cnt:e42639c66fe78808c04d8bedbb9ce84131562084;origin=https://github.com/jrh13/hol-light;visit=swh:1:snp:fe1b3b83e1dd2bde44d3698f723b25c949de2851;anchor=swh:1:rev:ab57c07ff0105fef75a9fcdd179eda0d26854ba3;path=/GL/gl.ml;lines=333-335} !*)
    $\vdash$ $\forall$p q. (*!\textbar!*)-- ((p --> q) && p --> q)
\end{lstlisting}
Moreover, we introduced some derived rules of the axiomatic system mimicking the behaviour in Gentzen's formalism of classical connectives, as in e.g.
\begin{lstlisting}[mathescape=true,escapeinside={(*!}{!*)}]
  GL_and_elim(*! \srclink{swh:1:cnt:e42639c66fe78808c04d8bedbb9ce84131562084;origin=https://github.com/jrh13/hol-light;visit=swh:1:snp:fe1b3b83e1dd2bde44d3698f723b25c949de2851;anchor=swh:1:rev:ab57c07ff0105fef75a9fcdd179eda0d26854ba3;path=/GL/gl.ml;lines=321-323} !*)
    $\vdash$ $\forall$p q r. (*!\textbar!*)-- (r --> p && q)
              $\Longrightarrow$ (*!\textbar!*)-- (r --> q) $\wedge$ (*!\textbar!*)-- (r --> p)
\end{lstlisting}
We had to prove about 120 such results of varying degrees of difficulty. We believe that this piece of code is well worth the effort of its development, for two main reasons to be considered -- along with the just mentioned fact that they provide a (not so) minimal set of internal lemmas which can be moved to different axiomatic calculi at, basically, no cost.

On the one hand, these lemmas simplify the subsequent formal proofs involving consistent lists of formulas since they let us work formally within the scope of \verb+|--+ so that we can rearrange subgoals according to their most useful equivalent form by applying the appropriate $\mathbb{GL}$-lemma(s).

On the other hand, giving formal proofs of these lemmas of the calculus $\mathbb{GL}$ has been important for checking how much our proof-assistant is ``friendly'' and efficient in performing this specific task.

As it is known, any axiomatic system fits very well an investigation involving a notion of \emph{theoremhood-as-tautology} for a specific logic, but its lack of naturalness w.r.t.~the practice of developing proper derivations makes it an unsatisfactory model for structural aspects of \emph{deducibility}. In more practical terms: developing a formal proof of a theorem in an axiomatic system \emph{by pencil and paper} can be a dull and uninformative task, even when dealing with trivial propositions.

We, therefore, left the proof search to the HOL Light toolbox as much as possible. Unfortunately, we have to express mixed feelings about the general experience. In most cases, relying on this specific proof assistant's automation tools did save our time and resources when trying to give a formal proof in $\mathbb{GL}$. Nevertheless, those automation tools did not turn out to be helpful at all in proving some $\mathbb{GL}$-lemmas. In those cases, we had to search for the specific instances of axioms from which deriving the lemmas,\footnote{The HOL Light tactics for first-order reasoning \texttt{MESON} and \texttt{METIS} were unable, for example, to instantiate autonomously the obvious middle formula for the transitivity of an implication, or even the specific formulas of a schema to apply to the goal in order to rewrite it.} so that interactive proving them had advantages as well as traditional instruments of everyday mathematicians.

To stress the general point: it is possible -- and valuable in general -- to rely on the resources of HOL Light to develop formal proofs both \emph{about} and \emph{within} an axiomatic calculus for a specific logic, in particular when the lemmas of the object system have relevance or practical utility for mechanising (meta-)results on it; however, these very resources -- and, as far as we can see, the tools of any other general proof assistant -- do not look particularly satisfactory for pursuing investigations on derivability within axiomatic systems.

\section{Modal completeness}\label{compl}
When dealing with normal modal logics, it is common to develop a proof of completeness w.r.t.~relational semantics by using the so-called `canonical model method'. This approach can be summarised as a standard construction of countermodels made of maximal consistent sets of formulas and an appropriate accessibility relation, according to e.g.~the textbooks \cite{popkorn1994first} and \cite{boolos1995logic}.

For $\mathbb{GL}$, we cannot pursue this strategy since the logic is not compact: maximal consistent sets are (in general) infinite objects, though the notion of derivability involves only a finite set of formulas. We cannot, therefore, reduce the semantic notion of (in)coherent set of formulas to the syntactic one of (in)consistent set of formulas: when extending a consistent set of formulas to a maximal consistent one, we might end up with a \emph{syntactically} consistent set that nevertheless cannot be \emph{semantically} satisfied.

Despite this, it is possible to achieve a completeness result by
\begin{enumerate}
\item identifying the relevant properties of maximal consistent sets of formulas; and
\item modifying the definitions so that those properties hold for specific consistent sets of formulas related to the formula to which we want to find a countermodel.
\end{enumerate}
That is the fundamental idea behind the proof in Boolos' monograph \cite[Ch.~5]{boolos1995logic}.
In that presentation, however, constructing a maximal consistent set from a simply consistent one is only proof-sketched and relies on a syntactic manipulation of formulas. By using HOL Light, we succeed in giving a detailed proof of completeness as directly as that by Boolos. Moreover, 
we can do that by carrying out in a \emph{very natural way} a tweaked Lindenbaum construction to extend consistent \emph{lists} to maximal consistent ones. This way, we succeed in preserving the standard Henkin-style completeness proofs, and, at the same time, we avoid the symbolic subtleties sketched in~\cite{boolos1995logic} that 
have the unpleasant consequence of making the formalised proof 
rather pedantic -- or even dull.

Furthermore, the proof of the main lemma \verb|EXTEND_MAXIMAL_CONSISTENT| is rather general and does not rely on any specific property of $\mathbb{GL}$:
our strategy suits all the other normal (mono)modal logics -- we only need to modify the subsequent definition of \verb|GL_STANDARD_REL| according to the specific system under consideration. Thus, we provide a way for formally establishing completeness à la Henkin \emph{and} the finite model property without resorting to filtrations~\cite{popkorn1994first} of canonical models for those systems.

\subsection{Maximal consistent lists}

Following the standard practice, we need to consider consistent finite sets of formulas for our proof of completeness. In principle, we can employ general sets of formulas in the formalisation. However, from the practical viewpoint, lists without repetitions are better suited since they are automatically finite
and we can easily manipulate them by structural recursion.
We define first the operation of finite conjunction of formulas in a list:\footnote{Notice that in this definition, we perform a case analysis where the singleton list is treated separately (i.e.~we have
\texttt{CONJLIST [p] = p}).  This is slightly uncomfortable in certain formal proof steps: in retrospect, we might have used a simpler version of this function. However, since this is a minor detail, we preferred not to change our code.}
\bigskip
\begin{lstlisting}[mathescape=true,escapeinside={(*!}{!*)}]
  let CONJLIST =(*! \srclink{swh:1:cnt:157e9cb72fd39cbce15433fd62df9fa3fb310ed3;origin=https://github.com/jrh13/hol-light;visit=swh:1:snp:fe1b3b83e1dd2bde44d3698f723b25c949de2851;anchor=swh:1:rev:ab57c07ff0105fef75a9fcdd179eda0d26854ba3;path=/GL/completeness.ml;lines=11-13} !*)
    new_recursive_definition list_RECURSION
    `CONJLIST [] = True $\wedge$
     ($\forall$p X. CONJLIST (CONS p X) =
            if X = [] then p else p && CONJLIST X)`;;
\end{lstlisting}

We prove some properties on lists of formulas and some $\mathbb{GL}$-lemmas involving \verb|CONJLIST|. 
These properties and lemma allow us to define the notion of \textbf{consistent list of formulas} and prove the main properties of this kind of objects:

\begin{lstlisting}[mathescape=true,escapeinside={(*!}{!*)}]
  let CONSISTENT = new_definition(*! \srclink{swh:1:cnt:157e9cb72fd39cbce15433fd62df9fa3fb310ed3;origin=https://github.com/jrh13/hol-light;visit=swh:1:snp:fe1b3b83e1dd2bde44d3698f723b25c949de2851;anchor=swh:1:rev:ab57c07ff0105fef75a9fcdd179eda0d26854ba3;path=/GL/completeness.ml;lines=149-150} !*)
    `CONSISTENT (l:form list) $\Longleftrightarrow$ $\neg$((*!\textbar!*)-- (Not (CONJLIST l)))`;;
\end{lstlisting}
In particular, we prove that:
\begin{itemize}
\item a consistent list cannot contain both \verb+p+ and $\texttt{Not}\;\mathtt{p}$ for any formula $\mathtt{p}$, nor \verb+False+;
\item for any consistent list $\mathtt{X}$ and formula $\mathtt{p}$, either $\mathtt{X}+\mathtt{p}$ is consistent, or $\mathtt{X}+\texttt{Not}\;\mathtt{p}$ is consistent,
where $+$ denotes the usual operation of appending an element to a list.
\end{itemize}
Our \textbf{maximal consistent lists} w.r.t.~a given formula $\mathtt{p}$ will be consistent lists that do not contain repetitions and that contain, for any subformula of $\mathtt{p}$, that very subformula or its negation:\footnote{Here we define the set of subformulas of $\mathtt{p}$ as the reflexive, transitive closure of the set of formulas on which the main connective of $\mathtt{p}$ operates: this way, the definition is simplified, and it is easier to establish standard properties of the set of subformulas employing general higher-order lemmas in HOL Light for the closure of a given relation.}

\begin{lstlisting}[mathescape=true,escapeinside={(*!}{!*)}]
  let MAXIMAL_CONSISTENT = new_definition(*! \srclink{swh:1:cnt:157e9cb72fd39cbce15433fd62df9fa3fb310ed3;origin=https://github.com/jrh13/hol-light;visit=swh:1:snp:fe1b3b83e1dd2bde44d3698f723b25c949de2851;anchor=swh:1:rev:ab57c07ff0105fef75a9fcdd179eda0d26854ba3;path=/GL/completeness.ml;lines=220-223} !*)
    `MAXIMAL_CONSISTENT p X $\Longleftrightarrow$
     CONSISTENT X $\wedge$ NOREPETITION X $\wedge$
     ($\forall$q. q SUBFORMULA p $\Longrightarrow$ MEM q X $\vee$ MEM (Not q) X)`;;
\end{lstlisting}
where \verb|X| is a list of formulas and \verb|MEM q X| is the membership relation for lists.
We then establish the main closure property (\verb+MAXIMAL_CONSISTENT_LEMMA+) of maximal consistent lists, namely closure under modus ponens
   \srclinkfoot{swh:1:cnt:157e9cb72fd39cbce15433fd62df9fa3fb310ed3;origin=https://github.com/jrh13/hol-light;visit=swh:1:snp:fe1b3b83e1dd2bde44d3698f723b25c949de2851;anchor=swh:1:rev:ab57c07ff0105fef75a9fcdd179eda0d26854ba3;path=/GL/completeness.ml;lines=225-243}.

After proving some further lemmas with practical utility -- in particular, the fact that any maximal consistent list behaves like a restricted bivalent evaluation for classical connectives 
-- we can finally define the ideal (type of counter)model we are interested in.

We define first the relation of \emph{subsentences} as
\begin{lstlisting}[mathescape=true,escapeinside={(*!}{!*)}]
   let SUBSENTENCE_RULES,SUBSENTENCE_INDUCT,(*! \srclink{swh:1:cnt:157e9cb72fd39cbce15433fd62df9fa3fb310ed3;origin=https://github.com/jrh13/hol-light;visit=swh:1:snp:fe1b3b83e1dd2bde44d3698f723b25c949de2851;anchor=swh:1:rev:ab57c07ff0105fef75a9fcdd179eda0d26854ba3;path=/GL/completeness.ml;lines=285-288} !*)
     SUBSENTENCE_CASES = new_inductive_definition
     `($\forall$p q. p SUBFORMULA q $\Longrightarrow$ p SUBSENTENCE q) $\wedge$
      ($\forall$p q. p SUBFORMULA q $\Longrightarrow$ Not p SUBSENTENCE q)`;;
 \end{lstlisting}
 
Next, given a formula \verb+p+, we take as ``standard'' -- and define the class \verb|GL_STANDARD_MODEL| -- those models consisting of:
\begin{enumerate}[label=C\arabic*. , ref=C\arabic*]
\item\label{1.} the set of maximal consistent lists w.r.t.~\verb|p| made of \emph{subsentences} of \verb|p| -- i.e.~its subformulas or their negations -- as possible worlds;
\item an accessibility relation \verb+R+ such that \begin{enumerate}[label=\alph*. , ref=\theenumi.\alph*]
    \item\label{2a.} it is irreflexive and transitive, and
    \item\label{2b.} for any subformula \verb|Box q| of \verb|p| and any world \verb|w|, \verb|Box q| is in \verb|w| iff, for any \verb|x| \verb|R|-accessible from \verb|w|, \verb|q| is in \verb|x|;
\end{enumerate} 
\item\label{3.} an atomic evaluation that gives value \verb|T| (true) to \verb|a| in \verb|w| iff \verb|a| is a subformula of \verb|p|.
\end{enumerate}


\noindent
The corresponding code is the following:
\begin{lstlisting}[mathescape=true,escapeinside={(*!}{!*)}]
  let GL_STANDARD_FRAME = new_definition(*! \srclink{swh:1:cnt:157e9cb72fd39cbce15433fd62df9fa3fb310ed3;origin=https://github.com/jrh13/hol-light;visit=swh:1:snp:fe1b3b83e1dd2bde44d3698f723b25c949de2851;anchor=swh:1:rev:ab57c07ff0105fef75a9fcdd179eda0d26854ba3;path=/GL/completeness.ml;lines=290-296} !*)
    `GL_STANDARD_FRAME p (W,R) $\Longleftrightarrow$
(*C1.  *) W = {w (*!\textbar!*) MAXIMAL_CONSISTENT p w $\wedge$
                   ($\forall$q. MEM q w $\Longrightarrow$ q SUBSENTENCE p)} $\wedge$
(*C2.a.*) ITF (W,R) $\wedge$
(*C2.b.*) ($\forall$q w. Box q SUBFORMULA p $\wedge$ w $\in$ W
            $\Longrightarrow$ (MEM (Box q) w $\Longleftrightarrow$ $\forall$x. R w x $\Longrightarrow$ MEM q x))`;;
\end{lstlisting}

\begin{lstlisting}[mathescape=true,escapeinside={(*!}{!*)}]
  let GL_STANDARD_MODEL = new_definition(*! \srclink{swh:1:cnt:157e9cb72fd39cbce15433fd62df9fa3fb310ed3;origin=https://github.com/jrh13/hol-light;visit=swh:1:snp:fe1b3b83e1dd2bde44d3698f723b25c949de2851;anchor=swh:1:rev:ab57c07ff0105fef75a9fcdd179eda0d26854ba3;path=/GL/completeness.ml;lines=310-313} !*)
    `GL_STANDARD_MODEL p (W,R) V $\Longleftrightarrow$
     GL_STANDARD_FRAME p (W,R) $\wedge$
(*C3.*)($\forall$a w. w $\in$ W
        $\Longrightarrow$ (V a w $\Longleftrightarrow$ MEM (Atom a) w $\wedge$ Atom a SUBFORMULA p))`;;
\end{lstlisting}

Notice that the conditions \ref{1.}, \ref{2b.} and \ref{3.} are very general and do not relate to the logic under consideration: they constitute a minimal set of conditions required for normal modal systems; on the contrary, the condition \ref{2a.} is specific to GL, and needs to be properly mirrrored at the syntactic level. That is the role played by the last two lines of the definition \verb+GL_STANDARD_REL+ discussed at the end of the next section.

\subsection{Maximal extensions}\label{sec:max}

What we have to do now is to show that the relation \verb|GL_STANDARD_MODEL| on the type of relational models is non-empty. We achieve this by constructing suitable maximal consistent lists of formulas from specific consistent ones.

Our original strategy differs from the presentation in e.g.~\cite{boolos1995logic} for being closer to the standard Lindenbaum construction commonly used to prove completeness results. By doing so, we have been able to circumvent both many technicalities in formalising the combinatorial argument sketched by Boolos in \cite[p.79]{boolos1995logic} \emph{and} the problem -- inherent to the Lindenbaum extension -- due to the non-compactness of the system, as we mentioned before.

The main lemma states then that, from any consistent list $\mathtt{X}$ of subsentences of a formula $\mathtt{p}$, we can construct a maximal consistent list of subsentences of $\mathtt{p}$ by extending (if necessary) $\mathtt{X}$:

\begin{lstlisting}[mathescape=true,escapeinside={(*!}{!*)}]
  EXTEND_MAXIMAL_CONSISTENT(*! \srclink{swh:1:cnt:157e9cb72fd39cbce15433fd62df9fa3fb310ed3;origin=https://github.com/jrh13/hol-light;visit=swh:1:snp:fe1b3b83e1dd2bde44d3698f723b25c949de2851;anchor=swh:1:rev:ab57c07ff0105fef75a9fcdd179eda0d26854ba3;path=/GL/completeness.ml;lines=485-545} !*)
    $\vdash$ $\forall$p X.
        CONSISTENT X $\wedge$
        ($\forall$q. MEM q X $\Longrightarrow$ q SUBSENTENCE p)
        $\Longrightarrow$ $\exists$M. MAXIMAL_CONSISTENT p M $\wedge$
                ($\forall$q. MEM q M $\Longrightarrow$ q SUBSENTENCE p) $\wedge$
                X SUBLIST M
\end{lstlisting}
where \verb|X SUBLIST M| denotes that each element of the list \verb|X| is an elment of the list \verb|M|.

The proof sketch is as follows: given a formula $\mathtt{p}$, we proceed in a step-by-step construction by iterating over the subformulas $\mathtt{q}$ of $\mathtt{p}$ not contained in \verb|X|.
At each step, we append to the list $\texttt{X}$ the subformula $\mathtt{q}$ -- if the resulting list is consistent -- or its negation $\texttt{Not}\; \mathtt{q}$ -- otherwise.

This way, we do not have to worry about the non-compactness of $\mathbb{GL}$ since we are working with finite objects -- the type \texttt{list} -- from the very beginning.


Henceforth, we see that -- under the assumption that $\mathtt{p}$ is not a $\mathbb{GL}$-lemma -- the set of possible worlds in \verb|GL_STANDARD_FRAME| w.r.t.~$\mathtt{p}$ is non-empty, as required by the definition of relational structures:

\begin{lstlisting}[mathescape=true,escapeinside={(*!}{!*)}]
  NONEMPTY_MAXIMAL_CONSISTENT(*! \srclink{swh:1:cnt:157e9cb72fd39cbce15433fd62df9fa3fb310ed3;origin=https://github.com/jrh13/hol-light;visit=swh:1:snp:fe1b3b83e1dd2bde44d3698f723b25c949de2851;anchor=swh:1:rev:ab57c07ff0105fef75a9fcdd179eda0d26854ba3;path=/GL/completeness.ml;lines=547-563} !*)
    $\vdash$ $\forall$p. $\neg$((*!\textbar!*)-- p)
          $\Longrightarrow$ $\exists$M. MAXIMAL_CONSISTENT p M $\wedge$
                  MEM (Not p) M $\wedge$
                  ($\forall$q. MEM q M $\Longrightarrow$ q SUBSENTENCE p)
\end{lstlisting}
Next, we have to define an $\mathtt{R}$ satisfying the conditions \ref{2a.} and \ref{2b.} for a
\verb|GL_STANDARD_FRAME|; the following does the job:

\begin{lstlisting}[mathescape=true,escapeinside={(*!}{!*)}]
  let GL_STANDARD_REL = new_definition(*! \srclink{swh:1:cnt:157e9cb72fd39cbce15433fd62df9fa3fb310ed3;origin=https://github.com/jrh13/hol-light;visit=swh:1:snp:fe1b3b83e1dd2bde44d3698f723b25c949de2851;anchor=swh:1:rev:ab57c07ff0105fef75a9fcdd179eda0d26854ba3;path=/GL/completeness.ml;lines=569-574} !*)
    `GL_STANDARD_REL p w x $\Longleftrightarrow$
     MAXIMAL_CONSISTENT p w $\wedge$
     ($\forall$q. MEM q w $\Longrightarrow$ q SUBSENTENCE p) $\wedge$
     MAXIMAL_CONSISTENT p x $\wedge$
     ($\forall$q. MEM q x $\Longrightarrow$ q SUBSENTENCE p) $\wedge$
     ($\forall$B. MEM (Box B) w $\Longrightarrow$ MEM (Box B) x $\wedge$ MEM B x) $\wedge$
     ($\exists$E. MEM (Box E) x $\wedge$ MEM (Not (Box E)) w)`;;
\end{lstlisting}
Notice that the last two lines of this definition assure that the condition \ref{2a.} is satisfied, i.e., that we considering irreflexive transitive frames.
Condition \ref{2b.} also needs to be satisfied: our \verb+ACCESSIBILITY_LEMMA+ assures that\footnote{Notice that we only need to prove the right-to-left direction of condition \ref{2b.}, since the converse is given by the second last requirement in the definition \verb+GL_STANDARD_FRAMES+. We can formally prove the former by reasoning within $\mathbb{GL}$, and using, in particular, the scheme $\mathsf{GL}$ and the derivability of $\Box p \rightarrow\Box p \wedge\Box\Box p$ (\verb+GL_dot_box+).}

\begin{lstlisting}[mathescape=true,escapeinside={(*!}{!*)}]
  ACCESSIBILITY_LEMMA(*! \srclink{swh:1:cnt:157e9cb72fd39cbce15433fd62df9fa3fb310ed3;origin=https://github.com/jrh13/hol-light;visit=swh:1:snp:fe1b3b83e1dd2bde44d3698f723b25c949de2851;anchor=swh:1:rev:ab57c07ff0105fef75a9fcdd179eda0d26854ba3;path=/GL/completeness.ml;lines=630-774} !*)
    $\vdash$ $\forall$p M w q.
        $\neg$((*!\textbar!*)-- p) $\wedge$
        MAXIMAL_CONSISTENT p M $\wedge$
        ($\forall$q. MEM q M $\Longrightarrow$ q SUBSENTENCE p) $\wedge$
        MAXIMAL_CONSISTENT p w $\wedge$
        ($\forall$q. MEM q w $\Longrightarrow$ q SUBSENTENCE p) $\wedge$
        MEM (Not p) M $\wedge$
        Box q SUBFORMULA p $\wedge$
        ($\forall$x. GL_STANDARD_REL p w x $\Longrightarrow$ MEM q x)
        $\Longrightarrow$ MEM (Box q) w,
\end{lstlisting}
Such a ``standard'' accessibility relation (together with the set of the specific maximal consistent lists we are dealing with) defines then a structure in \verb|ITF| with the required properties in order to satisfy the relation \lstinline+GL_STANDARD_FRAME+:

\begin{lstlisting}[mathescape=true,escapeinside={(*!}{!*)}]
  ITF_MAXIMAL_CONSISTENT(*! \srclink{swh:1:cnt:157e9cb72fd39cbce15433fd62df9fa3fb310ed3;origin=https://github.com/jrh13/hol-light;visit=swh:1:snp:fe1b3b83e1dd2bde44d3698f723b25c949de2851;anchor=swh:1:rev:ab57c07ff0105fef75a9fcdd179eda0d26854ba3;path=/GL/completeness.ml;lines=576-628} !*)
    $\vdash$ $\forall$p. $\neg$((*!\textbar!*)-- p)
           $\Longrightarrow$ ITF ({M | MAXIMAL_CONSISTENT p M $\wedge$
                         ($\forall$q. MEM q M $\Longrightarrow$ q SUBSENTENCE p)},
                    GL_STANDARD_REL p),
\end{lstlisting}

Notice that we might easily modify the formal proofs of conditions \ref{1.}, \ref{2b.} and \ref{3.} when dealing with different axiomatic systems, e.g.~$\mathbb{K},\mathbb{K}4,\mathbb{T},\mathbb{S}4,\mathbb{B},\mathbb{S}5$, as it happens at the informal level in Boolos \cite{boolos1995logic}.
In fact, for each of each systems, we would only have to modify the definition of the ``standard relation'' (in particular, the last two lines of our \verb+GL_STANDARD_REL+) and the parts of code in the proof of the accessibility lemma where we need to reason within the specific axiomatic calculus under investigation.
For each of these additional logics, a semantic condition on standard frames (line 4 of the definition \verb+GL_STANDARD_FRAME+) would then been defined formalising what \cite[Ch.~5]{boolos1995logic} reports on the topic.  

\subsection{Truth lemma and completeness}

For our ideal model, it remains to reduce the semantic relation of forcing to the more tractable one of membership to the specific world. More formally, we prove -- by induction on the complexity of the subformula $\mathtt{q}$ of $\mathtt{p}$ -- that if $\mathbb{GL}\not\vdash\mathtt{p}$, then for any world $\mathtt{w}$ of the standard model, $\mathtt{q}$ holds in $\mathtt{w}$ iff $\mathtt{q}$ is member of $\mathtt{w}$:\footnote{As before, this formal proof can be adapted to the other six modal systems mentioned at the end of the previous section.}

\begin{lstlisting}[mathescape=true,escapeinside={(*!}{!*)}]
  GL_truth_lemma(*! \srclink{swh:1:cnt:157e9cb72fd39cbce15433fd62df9fa3fb310ed3;origin=https://github.com/jrh13/hol-light;visit=swh:1:snp:fe1b3b83e1dd2bde44d3698f723b25c949de2851;anchor=swh:1:rev:ab57c07ff0105fef75a9fcdd179eda0d26854ba3;path=/GL/completeness.ml;lines=319-477} !*)
    $\vdash$ $\forall$W R p V q.
        $\neg$((*!\textbar!*)-- p) $\wedge$
        GL_STANDARD_MODEL p (W,R) V $\wedge$
        q SUBFORMULA p
        $\Longrightarrow$ $\forall$w. w $\in$ W $\Longrightarrow$ (MEM q w $\Longleftrightarrow$ holds (W,R) V q w),
\end{lstlisting}
Finally, we can prove the main result: if $\mathbb{GL}\not\vdash \mathtt{p}$, then the list $ [\mathtt{Not}\; \mathtt{p}]$ is consistent, and by applying \verb|EXTEND_MAXIMAL_CONSISTENT|, we obtain a maximal consistent list $\mathtt{X}$ w.r.t.~$\mathtt{p}$ that extends it, so that, by applying \verb|GL_truth_lemma|, we have that $\mathtt{X}\not\Vdash\mathtt{p}$ in our standard model. The corresponding formal proof reduces to the application of those previous results and the appropriate instantiations:

\begin{lstlisting}[mathescape=true,escapeinside={(*!}{!*)}]
  COMPLETENESS_THEOREM(*! \srclink{swh:1:cnt:157e9cb72fd39cbce15433fd62df9fa3fb310ed3;origin=https://github.com/jrh13/hol-light;visit=swh:1:snp:fe1b3b83e1dd2bde44d3698f723b25c949de2851;anchor=swh:1:rev:ab57c07ff0105fef75a9fcdd179eda0d26854ba3;path=/GL/completeness.ml;lines=830-840} !*)
    $\vdash$ $\forall$p. ITF (*!\textbar!*)= p $\Longrightarrow$ ((*!\textbar!*)-- p),
\end{lstlisting}
Notice that the family of frames \verb|ITF| is polymorphic, but, at this stage, our result holds only for frames on the domain \verb|form list|: the explicit type annotation would be
\begin{lstlisting}
    ITF : (form list->bool)#(form list->form list->bool)->bool
\end{lstlisting}
This is not an intrinsic limitation: the next section is devoted to generalising this theorem to frames on an arbitrary domain.



\subsection{Generalizing via bisimulation}
\label{sec:bisim}

As we stated before, our theorem~\verb|COMPLETENESS_THEOREM| provides the modal completeness for $\mathbb{GL}$ with respect to a semantics defined using models built on the type \verb|:form list|. This result would suffice to provide a (very non-optimal) bound on the complexity of $\mathbb{GL}:$ Testing the validity of a formula of size $n$ requires considering all models with cardinality $k$, for any $k\leq 2^{n}!$.

The same completeness result must also hold when considering models built on any infinite type. To obtain a formal proof of this fact, we need to establish a \emph{correspondence} between models built on different types. It is well-known that a good way to make rigorous such a correspondence is through the notion of \emph{bisimulation} \cite{blackburn}.

In our context, given two frames \verb|(W1,R1)| and \verb|(W2,R2)| sitting respectively on types \verb|:A| and \verb|:B|, each with an evaluation function \verb|V1| and \verb|V2|, a \textbf{bisimulation} is a binary relation \verb|Z:A->B->bool| that relates two worlds \verb|w1:A| and \verb|w2:B| when they can \emph{simulate} each other.  The formal definition is as follows:
\begin{lstlisting}[mathescape=true,escapeinside={(*!}{!*)}]
  BISIMIMULATION(*! \srclink{swh:1:cnt:e96b1bf9513c5683d53e971912903fc13d21b282;origin=https://github.com/jrh13/hol-light;visit=swh:1:snp:fe1b3b83e1dd2bde44d3698f723b25c949de2851;anchor=swh:1:rev:ab57c07ff0105fef75a9fcdd179eda0d26854ba3;path=/GL/modal.ml;lines=274-281} !*)
    $\vdash$ BISIMIMULATION (W1,R1,V1) (W2,R2,V2) Z $\Longleftrightarrow$
      ($\forall$w1:A w2:B.
       Z w1 w2
       $\Longrightarrow$ w1 $\in$ W1 $\wedge$ w2 $\in$ W2 $\wedge$
           ($\forall$a:string. V1 a w1 $\Longleftrightarrow$ V2 a w2) $\wedge$
           ($\forall$w1'. R1 w1 w1'
                  $\Longrightarrow$ $\exists$w2'. w2' $\in$ W2 $\wedge$ Z w1' w2' $\wedge$ R2 w2 w2') $\wedge$
           ($\forall$w2'. R2 w2 w2'
                  $\Longrightarrow$ $\exists$w1'. w1' $\in$ W1 $\wedge$ Z w1' w2' $\wedge$ R1 w1 w1'))
\end{lstlisting}
Then, we say that two worlds are \emph{bisimilar} if there exists a bisimulation between them:
\begin{lstlisting}[mathescape=true,escapeinside={(*!}{!*)}]
  let BISIMILAR = new_definition(*! \srclink{swh:1:cnt:e96b1bf9513c5683d53e971912903fc13d21b282;origin=https://github.com/jrh13/hol-light;visit=swh:1:snp:fe1b3b83e1dd2bde44d3698f723b25c949de2851;anchor=swh:1:rev:ab57c07ff0105fef75a9fcdd179eda0d26854ba3;path=/GL/modal.ml;lines=302-304} !*)
    `BISIMILAR (W1,R1,V1) (W2,R2,V2) (w1:A) (w2:B) $\Longleftrightarrow$
     $\exists$Z. BISIMIMULATION (W1,R1,V1) (W2,R2,V2) Z $\wedge$ Z w1 w2`;;
\end{lstlisting}
The key fact is that the semantic predicate \verb|holds| respects bisimilarity:


\begin{lstlisting}[mathescape=true,escapeinside={(*!}{!*)}]
  BISIMILAR_HOLDS(*! \srclink{swh:1:cnt:e96b1bf9513c5683d53e971912903fc13d21b282;origin=https://github.com/jrh13/hol-light;visit=swh:1:snp:fe1b3b83e1dd2bde44d3698f723b25c949de2851;anchor=swh:1:rev:ab57c07ff0105fef75a9fcdd179eda0d26854ba3;path=/GL/modal.ml;lines=311-315} !*)
    $\vdash$ $\forall$W1 R1 V1 W2 R2 V2 w1:A w2:B.
        BISIMILAR (W1,R1,V1) (W2,R2,V2) w1 w2
        $\Longrightarrow$ ($\forall$p. holds (W1,R1) V1 p w1 $\Longleftrightarrow$
                 holds (W2,R2) V2 p w2)
\end{lstlisting}
From this, we can prove that bisimilarity preserves validity.  The precise statements are the following:
\begin{lstlisting}[mathescape=true,escapeinside={(*!}{!*)}]
  BISIMILAR_HOLDS_IN(*! \srclink{swh:1:cnt:e96b1bf9513c5683d53e971912903fc13d21b282;origin=https://github.com/jrh13/hol-light;visit=swh:1:snp:fe1b3b83e1dd2bde44d3698f723b25c949de2851;anchor=swh:1:rev:ab57c07ff0105fef75a9fcdd179eda0d26854ba3;path=/GL/modal.ml;lines=317-321} !*)
    $\vdash$ $\forall$W1 R1 W2 R2.
         ($\forall$V1 w1:A.
            $\exists$V2 w2:B. BISIMILAR (W1,R1,V1) (W2,R2,V2) w1 w2)
         $\Longrightarrow$ ($\forall$p. holds_in (W2,R2) p $\Longrightarrow$ holds_in (W1,R1) p)
\end{lstlisting}
\begin{lstlisting}[mathescape=true,escapeinside={(*!}{!*)}]
  BISIMILAR_VALID(*! \srclink{swh:1:cnt:e96b1bf9513c5683d53e971912903fc13d21b282;origin=https://github.com/jrh13/hol-light;visit=swh:1:snp:fe1b3b83e1dd2bde44d3698f723b25c949de2851;anchor=swh:1:rev:ab57c07ff0105fef75a9fcdd179eda0d26854ba3;path=/GL/modal.ml;lines=323-332} !*)
    $\vdash$ $\forall$L1 L2.
        ($\forall$W1 R1 V1 w1:A.
          L1 (W1,R1) $\wedge$ w1 $\in$ W1
          $\Longrightarrow$ $\exists$W2 R2 V2 w2:B.
                L2 (W2,R2) $\wedge$
                BISIMILAR (W1,R1,V1) (W2,R2,V2) w1 w2)
        $\Longrightarrow$ ($\forall$p. L2 (*!\textbar!*)= p $\Longrightarrow$ L1 (*!\textbar!*)= p)
\end{lstlisting}
In the last theorem, recall that the statement \verb|L(W,R)| means that \verb|(W R)| is a frame in the class of frames \verb|L|.

Finally, we can explicitly define a bisimulation between \verb|ITF|-models on the type \verb|:form list| and on any infinite type \verb|:A|.
From this, it follows at once the desired generalization of completeness for $\mathbb{GL}$:
\begin{lstlisting}[mathescape=true,escapeinside={(*!}{!*)}]
  COMPLETENESS_THEOREM_GEN(*! \srclink{swh:1:cnt:157e9cb72fd39cbce15433fd62df9fa3fb310ed3;origin=https://github.com/jrh13/hol-light;visit=swh:1:snp:fe1b3b83e1dd2bde44d3698f723b25c949de2851;anchor=swh:1:rev:ab57c07ff0105fef75a9fcdd179eda0d26854ba3;path=/GL/completeness.ml;lines=846-892} !*)
    $\vdash$ $\forall$p. INFINITE (:A) $\wedge$ ITF:(A->bool)#(A->A->bool)->bool (*!\textbar!*)= p
          $\Longrightarrow$ (*!\textbar!*)-- p
\end{lstlisting}


Furthermore, from the proof that the relation
\begin{lstlisting}[mathescape=true,escapeinside={(*!}{!*)}]
$\lambda$w1 w2. MAXIMAL_CONSISTENT p w1 $\wedge$
        ($\forall$q. MEM q w1 $\Longrightarrow$ q SUBSENTENCE p) $\wedge$
        w2 $\in$ GL_STDWORLDS p $\wedge$
        set_of_list w1 = w2
\end{lstlisting}
defines a bisimulation between the \texttt{ITF}-standard model based on maximal consistent \emph{lists} of formulas and the model based on corresponding \emph{sets} of formulas, we obtain the traditional version of modal completeness, corresponding to theorem \verb|GL_COUNTERMODEL_FINITE_SETS| \srclinkfoot{swh:1:cnt:157e9cb72fd39cbce15433fd62df9fa3fb310ed3;origin=https://github.com/jrh13/hol-light;visit=swh:1:snp:fe1b3b83e1dd2bde44d3698f723b25c949de2851;anchor=swh:1:rev:ab57c07ff0105fef75a9fcdd179eda0d26854ba3;path=/GL/completeness.ml;lines=1052-1073} in our code.
That, in turn, would enhance the complexity measure of derivability in $\mathbb{GL}$ as one would expect: now, in order to check whether a formula with size $n$ is a theorem of G\"odel-L\"ob logic, one may forget about the order of the subsentences, and consider all $\mathtt{ITF}$-models with cardinality $k$, for any $k\leq 2^{n}$.

\section{Decidability via proof theory}\label{prooftheorygl}
By using our \verb|EXTEND_MAXIMAL_CONSISTENT| lemma, we succeeded in giving a rather neat mechanised proof of completeness w.r.t.\ relational semantics for $\mathbb{GL}$.\footnote{For the completeness w.r.t.~transitive Noetherian frames, it is common -- see the textbooks~\cite{boolos1995logic,popkorn1994first} -- to reason on irreflexive transitive finite structures and derive the result as a corollary of completeness w.r.t.~the latter class.} Since the relational countermodel we construct is finite, we can safely claim\footnote{After Harrop \cite{harrop_1958}.} that the finite model property holds for $\mathbb{GL}$.

As an immediate corollary, we have that theoremhood in $\mathbb{GL}$ is decidable, and, in principle, we could implement a decision procedure for that logic in OCaml. Our theorem \verb|GL_COUNTERMODEL_FINITE_SETS| would also provide an explicit bound on the complexity of that task.

A naive approach to the implementation would proceed as follows.

We define the tactic \verb|NAIVE_GL_TAC| and its associated rule \verb|NAIVE_GL_RULE| that perform the following steps:
\begin{enumerate}
\item Apply the completeness theorem w.r.t.\ finite sets;
\item Unfold some definitions;
\item Try to solve the resulting \emph{semantic} problem using first-order reasoning.
\end{enumerate}
Here is the corresponding code.

\begin{lstlisting}[mathescape=true,escapeinside={(*!}{!*)}]
  let NAIVE_GL_TAC : tactic =(*! \srclink{swh:1:cnt:157e9cb72fd39cbce15433fd62df9fa3fb310ed3;origin=https://github.com/jrh13/hol-light;visit=swh:1:snp:fe1b3b83e1dd2bde44d3698f723b25c949de2851;anchor=swh:1:rev:ab57c07ff0105fef75a9fcdd179eda0d26854ba3;path=/GL/completeness.ml;lines=898-902} !*)
    MATCH_MP_TAC GL_COUNTERMODEL_FINITE_SETS THEN
    REWRITE_TAC[valid; FORALL_PAIR_THM; holds_in; holds;
                ITF; GSYM MEMBER_NOT_EMPTY] THEN
    MESON_TAC[];;

  let NAIVE_GL_RULE tm = prove(tm, REPEAT GEN_TAC THEN GL_TAC);;
\end{lstlisting}
The above strategy can already prove some lemmas common to normal modal logics automatically but require some effort when derived in an axiomatic system. As an example, consider the following $\mathbb{GL}$-lemma:
\begin{lstlisting}[mathescape=true,escapeinside={(*!}{!*)}]
  GL_box_iff_th(*! \srclink{swh:1:cnt:e42639c66fe78808c04d8bedbb9ce84131562084;origin=https://github.com/jrh13/hol-light;visit=swh:1:snp:fe1b3b83e1dd2bde44d3698f723b25c949de2851;anchor=swh:1:rev:ab57c07ff0105fef75a9fcdd179eda0d26854ba3;path=/GL/gl.ml;lines=900-911} !*)
    $\vdash$ $\forall$p q. (*!\textbar!*)-- (Box (p <-> q) --> (Box p <-> Box q))
\end{lstlisting}

When developing a proof of it within the axiomatic calculus, we need to ``help'' HOL Light by instantiating several further $\mathbb{GL}$-lemmas so that the resulting proof script consists of ten lines of code.
On the contrary, our rule is able to check it in a few steps:
\begin{lstlisting}[mathescape=true,escapeinside={(*!}{!*)}]
  # NAIVE_GL_RULE
      `!p q. (*!\textbar!*)-- (Box (p <-> q) --> (Box p <-> Box q))`;;
  0..0..1..6..11..19..32..solved at 39
  0..0..1..6..11..19..32..solved at 39
  val it : thm =
    (*!\textbar!*)- !p q. (*!\textbar!*)-- (Box (p <-> q) --> (Box p <-> Box q))
\end{lstlisting}
In spite of this, the automation offered by \verb|MESON| tactic is often unsuccessful, even on trivial lemmas. For instance, the previous procedure is not even able to prove the basic instance Gödel-Löb scheme
$$\mathbb{GL} \vdash \Box(\Box\bot \longrightarrow \bot) \longrightarrow \Box\bot\,.$$

This suggests that \verb|NAIVE_GL_TAC| is based on a very inadequate strategy.

A better approach would consist of introducing an OCaml bound function on the size of the frames on which the validity of a formula $A$ has to be checked, according to the cardinality remarks we made at the end of Section \ref{sec:bisim}.

That is, for sure, a doable way to solve the task, but we were looking for a more principled and modern approach.

That is where structural proof theory has come to the rescue.

\subsection{Bits of structural proof theory}\label{sec:bits}
In very abstract terms, a deductive system consists of a set of starting formal expressions together with inference rules. Its principal aim is to find proofs of valid expressions w.r.t.~a given logic L. A proof (or derivation) in a deductive system is obtained by application of the inference rules to starting expressions, followed by further application of the inference rules to the conclusion, and so on, recursively. A theorem (or lemma) in such a system is the formal expression obtained after a finite run of the procedure just sketched.

Such a definition captures the system $\mathbb{GL}$, and the derivability relation formalised in HOL Light by the predicate \verb+|--+ of Section \ref{axiom}. Our previous remarks on $\mathbb{GL}$-lemmas make explicit that this kind of deductive system has shortcomings that its computerisation per se cannot overcome.

The proof-theoretic paradigm behind $\mathbb{GL}$ and, more generally, axiomatic calculi could be called \emph{synthetic}: proof search in such systems is not guided by the components of the formula one wishes to prove. The human prover and the proof assistant dealing with a formal derivability relation have to guess both the correct instances of the axiom schemas and the correct application order of inference rules required in the proof.

A better paradigm is provided by Gentzen's sequent calculi, introduced first in \cite{gentzen1935untersuchungen1,gentzen1935untersuchungen2}. That work marks the advent of structural proof theory and the definite shift from investigations in synthetic deductive systems to \emph{analytic} ones.
Gentzen's original calculi have been further refined by Ketonen \cite{ketonen1945untersuchungen} and Kleene \cite{kleene1952permutability} into the so-called $\mathsf{G3}$-style systems.

In those systems, a \textbf{sequent} is a formal expression with shape $$\Gamma\Rightarrow\Delta,$$ where $\Gamma,\Delta$ are finite multisets -- i.e.~finite lists modulo permutations -- of formulas of a given language. The symbol $\Rightarrow$ reflects the deducibility relation at the meta-level in the object language. $\Gamma$ is called the \textbf{antecedent} of the sequent; $\Delta$ is its \textbf{consequent}.

A derivation in a $\mathsf{G3}$-style sequent calculus is a finite rooted tree labelled with sequents such that:
\begin{itemize}
\item its leaves are labelled by \textbf{initial sequents} (the starting formal expressions of the abstract deductive system);
\item its intermediate nodes are labelled by sequents obtained from the sequent(s) labelling the node(s) directly above by a correct application of an inference rule of the calculus;
\item its root is the conclusion of the derivation, and it is called the \textbf{end-sequent}.
\end{itemize}

Figure \ref{fig:G3cp} summarises the calculus $\mathsf{G3cp}$ for classical propositional logic. For each rule, one distinguishes:
\begin{itemize}
\item its \textbf{main formula}, which is the formula occurring in the conclusion and containing the logical connective naming the rule;
\item its \textbf{active formulas}, which are the formulas occurring in the premise(s) of the rule;
\item its \textbf{context}, which consists of the formulas occurring in the premise(s) \emph{and} the conclusion, untouched by the rule.
\end{itemize}

\begin{figure}[h!]

\begin{center}
\begin{small}
\hrulefill

\begin{tabular}{l l}

& \\

\textbf{Initial sequents:}\\
& \\
$p,\Gamma\Rightarrow\Delta,p$\\
& \\
& \\
\textbf{Propositional rules:}\\
& \\
\AxiomC{$\,$}\RightLabel{$\mathsmaller{\mathcal{L}\bot}$}\UnaryInfC{$\bot,\Gamma\Rightarrow\Delta$}\DisplayProof\\
& \\
\AxiomC{$A,B,\Gamma\Rightarrow\Delta$}\RightLabel{$\mathsmaller{\mathcal{L}\wedge}$}\UnaryInfC{$A\wedge B,\Gamma\Rightarrow\Delta$}\DisplayProof & \AxiomC{$\Gamma\Rightarrow\Delta,A$}\AxiomC{$\Gamma\Rightarrow\Delta,B$}\RightLabel{$\mathsmaller{\mathcal{R}\wedge}$}\BinaryInfC{$\Gamma\Rightarrow\Delta,A\wedge B$}\DisplayProof \\
& \\
\AxiomC{$A,\Gamma\Rightarrow\Delta$}\AxiomC{$B,\Gamma\Rightarrow\Delta$}\RightLabel{$\mathsmaller{\mathcal{L}\vee}$}\BinaryInfC{$A\vee B,\Gamma\Rightarrow\Delta$}\DisplayProof & \AxiomC{$\Gamma\Rightarrow\Delta,A,B$}\RightLabel{$\mathsmaller{\mathcal{R}\vee}$}\UnaryInfC{$\Gamma\Rightarrow\Delta,A\vee B$}\DisplayProof \\
& \\
\AxiomC{$\Gamma\Rightarrow\Delta,A$}\RightLabel{$\mathsmaller{\mathcal{L}\neg}$}\UnaryInfC{$\neg A,\Gamma\Rightarrow\Delta$}\DisplayProof & \AxiomC{$A,\Gamma\Rightarrow\Delta$}\RightLabel{$\mathsmaller{\mathcal{R}\neg}$}\UnaryInfC{$\Gamma\Rightarrow\Delta,\neg A$}\DisplayProof \\
& \\
\AxiomC{$\Gamma\Rightarrow\Delta,A$}\AxiomC{$B,\Gamma\Rightarrow\Delta$}\RightLabel{$\mathsmaller{\mathcal{L}\rightarrow}$}\BinaryInfC{$A\rightarrow B,\Gamma\Rightarrow\Delta$}\DisplayProof & \AxiomC{$A,\Gamma\Rightarrow\Delta,B$}\RightLabel{$\mathsmaller{\mathcal{R}\rightarrow}$}\UnaryInfC{$\Gamma\Rightarrow\Delta,A\rightarrow B$}\DisplayProof \\
& \\

\end{tabular}

\hrulefill

\end{small}
\end{center}
\caption{Rules of the calculus $\mathsf{G3cp}$}
\label{fig:G3cp}
\end{figure}

$\mathsf{G3}$-style systems are the best available option for (efficiently) automating decision procedures: once an adequate -- i.e.~sound and complete -- $\mathsf{G3}$-calculus for a given logic L has been defined, in order to decide whether a formula is a theorem of L it suffices to start a \emph{root-first} proof search of that exact formula in the related $\mathsf{G3}$-calculus.

This is so because, by design, good $\mathsf{G3}$-style systems satisfy the following desiderata:
\begin{enumerate}
\item \textbf{Analyticity:} Each formula occurring in a derivation is a subformula of the formulas occurring lower in the derivation branch. This means that \emph{no guesses are required} to the prover when developing a formal proof in the $\mathsf{G3}$-calculus;
\item \textbf{Invertibility of all the rules:} For each rule of the system, the derivability of the conclusion implies the derivability of the premise(s). \emph{This property avoids backtracking} on the proof search. It also means that at each step of the proof search procedure \emph{no bit of information gets lost}, so that the construction of \emph{one} derivation tree is enough to decide the derivability of a sequent;
\item \textbf{Termination:} Each proof search must come to an end. If the procedure's final state generates a derivation, the end-sequent is a theorem; otherwise, it is generally possible to extract a \emph{refutation} of the sequent from the failed proof search.\footnote{Notice, however, that sometimes the invertibility of a rule could break termination of the proof search, as witnessed by $\mathcal{L}\rightarrow$ in $\mathsf{G3ip}$ for intuitionistic propositional logic \cite{troelstra2000basic}.}
\end{enumerate}




Satisfaction of all those desiderata by a pure Gentzen-style sequent calculus has been considered for long almost a mirage for non-classical logics.
Times have changed with the advent of \emph{internalisation} techniques of semantic notions in sequent calculi for non-classical logics.

The starting point of that perspective is still the basic $\mathsf{G3}$-paradigm, but the formalism of the sequent system is extended either by
\begin{itemize}
\item enriching the language of the calculus itself (\textbf{explicit internalisation}); or by
\item enriching the structure of sequents (\textbf{implicit internalisation}).
\end{itemize}

The implicit approach adds structural connectives to sequents other than `$\Rightarrow$' and commas.\footnote{
Refer to e.g.\ Avron\cite{avron1996method}, Mints \cite{mints1974sistemy,mints1992short}, and Pottinger \cite{pottinger1983uniform}, as well as Restall \cite{restall2005proofnets}, Kurokawa \cite{kurokawa2013hypersequent}, Marin and Stra{\ss}burger \cite{marin2014label}. 
For its direct generalisation by nested sequents refer to e.g.\ Kashima \cite{kashima1994cut}, Br\"unnler \cite{brunnler2009deep}, as well as Poggiolesi \cite{poggiolesi2009method,poggiolesi2010gentzen,poggiolesi2016natural} and Olivetti and Pozzato \cite{olivetti2015standard}.
}
Most $\mathsf{G}3$-style calculi obtained this way provide, in general, very efficient decision procedures for the related logics. However, they are sometimes rather hard to design and might lack an additional and highly valuable desideratum of modularity.

The explicit approach reverses the situation. Explicit internalisation uses specific items to represent semantic elements; the formulas of the basic language are then \textbf{labelled} by those items and have shape e.g.~$x : A$. That expression formalises the forcing relation we rephrased in Section \ref{sec:frames-GL}, and classical propositional rules operate within the scope of labels. Furthermore, to handle modal operators, the antecedent of any sequent may now contain \textbf{relational atoms} of shape $xRy$, or any other expression borrowed from the semantics for the logic under investigation.

The rules for the modalities formalise the forcing condition for each modal connective. For instance, the rules in Figure \ref{fig:g3k} define the labelled sequent calculus $\mathsf{G3K}$.

\begin{figure}[h!]

\begin{center}
\begin{small}
\hrulefill

\begin{tabular}{l l}

& \\

\textbf{Initial sequents:}\\
& \\
$x:p,\Gamma\Rightarrow\Delta,x:p$\\
& \\
& \\
\textbf{Propositional rules:}\\
& \\
\AxiomC{$\,$}\RightLabel{$\mathsmaller{\mathcal{L}\bot}$}\UnaryInfC{$x:\bot,\Gamma\Rightarrow\Delta$}\DisplayProof\\
& \\
\AxiomC{$x:A,x:B,\Gamma\Rightarrow\Delta$}\RightLabel{$\mathsmaller{\mathcal{L}\wedge}$}\UnaryInfC{$x:A\wedge B,\Gamma\Rightarrow\Delta$}\DisplayProof & \AxiomC{$\Gamma\Rightarrow\Delta,x:A$}\AxiomC{$\Gamma\Rightarrow\Delta,x:B$}\RightLabel{$\mathsmaller{\mathcal{R}\wedge}$}\BinaryInfC{$\Gamma\Rightarrow\Delta,x:A\wedge B$}\DisplayProof \\
& \\
\AxiomC{$x:A,\Gamma\Rightarrow\Delta$}\AxiomC{$x:B,\Gamma\Rightarrow\Delta$}\RightLabel{$\mathsmaller{\mathcal{L}\vee}$}\BinaryInfC{$x:A\vee B,\Gamma\Rightarrow\Delta$}\DisplayProof & \AxiomC{$\Gamma\Rightarrow\Delta,x:A,x:B$}\RightLabel{$\mathsmaller{\mathcal{R}\vee}$}\UnaryInfC{$\Gamma\Rightarrow\Delta,x:A\vee B$}\DisplayProof \\
& \\
\AxiomC{$\Gamma\Rightarrow\Delta,x:A$}\RightLabel{$\mathsmaller{\mathcal{L}\neg}$}\UnaryInfC{$x:\neg A,\Gamma\Rightarrow\Delta$}\DisplayProof & \AxiomC{$x:A,\Gamma\Rightarrow\Delta$}\RightLabel{$\mathsmaller{\mathcal{R}\neg}$}\UnaryInfC{$\Gamma\Rightarrow\Delta,x:\neg A$}\DisplayProof \\
& \\
\AxiomC{$\Gamma\Rightarrow\Delta,x:A$}\AxiomC{$x:B,\Gamma\Rightarrow\Delta$}\RightLabel{$\mathsmaller{\mathcal{L}\rightarrow}$}\BinaryInfC{$x:A\rightarrow B,\Gamma\Rightarrow\Delta$}\DisplayProof & \AxiomC{$x:A,\Gamma\Rightarrow\Delta,x:B$}\RightLabel{$\mathsmaller{\mathcal{R}\rightarrow}$}\UnaryInfC{$\Gamma\Rightarrow\Delta,x:A\rightarrow B$}\DisplayProof \\
& \\
& \\
\textbf{Modal rules:}\\
& \\
\AxiomC{$y:A,xRy,x:\Box A,\Gamma\Rightarrow\Delta$}\RightLabel{$\mathsmaller{\mathcal{L}\Box}$}\UnaryInfC{$xRy,x:\Box A,\Gamma\Rightarrow\Delta$}\DisplayProof & \AxiomC{$xRy,\Gamma\Rightarrow\Delta,y:A$}\RightLabel{$\mathsmaller{\mathcal{R}\Box_{(y!)}}$}\UnaryInfC{$\Gamma\Rightarrow\Delta,x:\Box A$}\DisplayProof \\
& \\

\end{tabular}

where the annotation $(y!)$ in $\mathcal{R}\Box$ states that the label $y$ does not occur in $\Gamma,\Delta$.

\medskip

\hrulefill

\end{small}
\end{center}
\caption{Rules of the calculus $\mathsf{G3K}$}
\label{fig:g3k}
\end{figure}

Extensions of the minimal normal modal logic K are obtained by rules for relational atoms, formalising the characteristic properties of each specific extension. For instance, the system $\mathsf{G3K4}$ for the logic K4 is defined by adding to $\mathsf{G3K}$ the rule
\begin{prooftree}\AxiomC{$xRz,xRy,yRz,\Gamma\Rightarrow\Delta$}\RightLabel{\small{\textit{Trans}}}\UnaryInfC{$xRy,yRz,\Gamma\Rightarrow\Delta$}
\end{prooftree}

Labelled sequent calculi do satisfy, in general, the basic desiderata of $\mathsf{G3}$-style systems and are rather modular.\footnote{
Their origin dates back to Kanger \cite{kanger1957provability}, followed by the refinements in Kripke \cite{kripke}, Fitting \cite{fitting2013proof}, and Gabbay \cite{gabbay1996labelled}. However, labelled sequent calculi have been established as a well-structured methodology after Negri \cite{negri2005proof}. Extensions, refinements, and further results have since been obtained for many logics. Refer to e.g.~\cite{negri2017proof, negri2021proof}, Poggiolesi \cite{poggiolesi2016natural}, \cite{dalmonte2018non}.
} However, since analyticity holds in a less strict version than the subformula principle mentioned in the previous definition, termination of proof search is sometimes hard to prove and, in many cases, produces complexity results far from optimal.


In the present paper, internalisation is considered only in its explicit version for some syntactic economy: we adopted Negri's labelled sequent calculus \cite{negri2005proof} to achieve our goal of implementing a decision algorithm for GL.

\medskip

Before proceeding, let us clarify the methodology behind labelled sequents for normal modal logics.

For those logics, labelled sequent calculi are based on a language $\mathcal{L}_{LS}$ which extends $\mathcal{L}$ with a set of world labels and a set of relational atoms of the form $xRy$ for world labels $x,y$. Formulas of $\mathcal{L}_{LS}$ have only two possible forms $$x:A\qquad\textrm{or}\qquad xRy,$$ where $A$ is a formula of $\mathcal{L}$.

Sequents are now pairs $\Gamma\Rightarrow\Delta$ of multisets of formulas of $\mathcal{L}_{LS}$.


Accordingly, the rules for \emph{all} logical operators are based on the inductive definition of the forcing relation between worlds and formulas of $\mathcal{L}$, here denoted by $x:A$. In particular, the forcing condition for the $\Box$ modality $$x\Vdash\Box A\qquad\textrm{iff}\qquad\textrm{for all}\, y,\,\textrm{if}\, xRy,\,\textrm{then}\, y\Vdash A$$ determines the following rules:

\medskip
\begin{center}
\AxiomC{$xRy,\Gamma\Rightarrow\Delta,y:A$}\RightLabel{$\mathsmaller{\mathcal{R}\Box_{(y!)}}$}\UnaryInfC{$\Gamma\Rightarrow\Delta,x:\Box A$}\DisplayProof
$\qquad\qquad$
\AxiomC{$y:A,xRy,x:\Box A,\Gamma\Rightarrow\Delta$}\RightLabel{$\mathsmaller{\mathcal{L}\Box}$}\UnaryInfC{$xRy,x:\Box A,\Gamma\Rightarrow\Delta$}\DisplayProof
\end{center}
\medskip
with  the side condition for $\mathcal{R}\Box$ that $y$ does not occur in $\Gamma,\Delta$.

This is the very general framework behind the labelled sequent calculus $\mathsf{G3K}$ for the basic normal modal logic $\mathbb{K}$ we defined in Figure \ref{fig:g3k}. The results presented by Negri and von Plato in \cite[Ch.~6]{negri2008structural} assure that any extension of $\mathbb{K}$ that is semantically characterised by (co)geometric frame conditions\footnote{Recall that a frame condition is said to be geometric \cite{negri2011proof} when it has shape $\forall\Vec{x}(\varphi\rightarrow\psi),$ where $\varphi,\psi$ are first-order formulas that do not contain universal quantifiers or implications. This means that the semantics behind the logical system is a geometric theory in the sense explained in \cite{dyckhoff_negri_2015}.} can also be captured by an extension of $\mathsf{G3K}$ with appropriate (co)geometric rules.

Any of such extensions will keep the good structural properties of $\mathsf{G3cp}$. Most relevantly, proving termination of the proof search in these calculi provides a neat proof of decidability for the corresponding logics, which allows the direct construction of a countermodel for a given formula generating a failed proof search.
In many cases, such a proof of decidability is obtained as follows: because of the structural correspondence between the proof tree generated during a backward proof search and the relational frames at the semantic level, proving the finite model property (hence, decidability) for a given logic reduces to proving that the such a root-first search generates only a finite number of new labels; in general, a specific proof search strategy needs to be defined for each system under consideration, in order to prevent looping interactions of (some of) the rules of the given calculus. A detailed account of the methodology is given in Garg et al.~\cite{6280450}.   

\section{Implementing \texorpdfstring{$\mathsf{G3KGL}$}{G3KGL}}\label{decision}
The frame conditions characterising GL -- i.e.~Noetherianity and transitivity, or, equivalently, irreflexivity, transitivity and finiteness -- cannot be expressed by a (co)geometric implication since finiteness and Noetherianity are intrinsically second order.

Therefore it is not possible to define a labelled sequent calculus for GL by simply extending $\mathsf{G3K}$ with naive semantic rules.

Nevertheless, it is still possible to internalise the possible world semantics by relying on a modified definition of the forcing relation -- valid for $\mathtt{ITF}$ models -- in which the standard condition for $\Box$ is substituted by the following
\begin{equation}\label{eq:gl}
    x\Vdash\Box A\qquad\textrm{iff}\qquad\textrm{for all}\, y,\,\textrm{if}\;xRy\;\textrm{and}\;y\Vdash\Box A,\,\textrm{then}\;y\Vdash A.
\end{equation}

Following Negri \cite{negri2005proof}, this suggests modifying the $\mathcal{R}\Box$ rule according to the right-to-left direction of (\ref{eq:gl}).

The resulting labelled sequent calculus $\mathsf{G3KGL}$ is then characterised by the initial sequents, the propositional rules and the $\mathcal{L}\Box$ in Figure \ref{fig:g3k} together with the rules in Figure \ref{fig:g3kgl}.

\begin{figure}[h!]

\begin{center}

\begin{small}

\hrulefill

\begin{tabular}{l l}
& \\

& \AxiomC{$xRy,y:\Box A,\Gamma\Rightarrow\Delta,y:A$}\RightLabel{$\mathsmaller{\mathcal{R}\Box^{\textit{L\"ob}}_{(y!)}}$}\UnaryInfC{$\Gamma\Rightarrow\Delta,x:\Box A$}\DisplayProof \\
& \\
& \\
\AxiomC{$\,$}\RightLabel{$\mathsmaller{\textit{Irref}}$}\UnaryInfC{$xRx,\Gamma\Rightarrow\Delta$}\DisplayProof$\qquad\qquad$ & \AxiomC{$xRz,xRy,yRz,\Gamma\Rightarrow\Delta$}\RightLabel{$\mathsmaller{\textit{Trans}}$}\UnaryInfC{$xRy,yRz,\Gamma\Rightarrow\Delta$}\DisplayProof

\end{tabular}

\hrulefill

\end{small}
\end{center}

\caption{Additional rules for the calculus $\mathsf{G3KGL}$}\label{fig:g3kgl}
\end{figure}

The rule $\mathcal{R}\Box^{\textit{L\"ob}}$ obeys to the side condition of not being $y$ free in $\Gamma,\Delta$, in line with the universal quantifier used in the forcing condition.

In Negri \cite{negri2014proofs}, it is proven that $\mathsf{G3KGL}$ has good structural properties. Moreover, in the same paper, it is easily shown that the proof search in this calculus is terminating. Its failure allows to construct a countermodel for the formula at the root of the derivation tree: it suffices to consider the top sequent of an open branch, and assume that all the labelled formulas and relational atoms in its antecedent are true, while all the labelled formulas in its consequent are false. Termination is assured by imposing saturation conditions on sequents that might prevent the application of useless rules, according to the Sch\"utte-Takeuti-Tait reduction for classical $\mathsf{G3}$ calculi. The reader is referred to Negri \cite{negri2014proofs} for a detailed description of the proof search algorithm for $\mathsf{G3KGL}$.\\
This way, a syntactic proof of decidability for GL is obtained.


\subsection{How to use \texorpdfstring{$\mathsf{G3KGL}$}{G3KGL}}
It is not hard to see how to use both our formalisation of modal completeness and the already known proof theory for $\mathsf{G3KGL}$ to the aim of implementing a decision algorithm in HOL Light for GL: our predicate \verb|holds (W,R) V A x| corresponds precisely to the labelled formula $x:A$. Thus we have three different ways of expressing the fact that a world $x$ forces $A$ in a given model $\langle W,R,V\rangle$:

\medskip

\begin{center}
\begin{tabular}{l | l}
\textsc{Semantic notation} & $x\Vdash A$ \\ \hline
\textsc{Labelled sequent calculus notation} & $x:A$ \\ \hline
\textsc{HOL Light notation} & \verb|holds (W,R) V A x|
\end{tabular}
\end{center}

\medskip

This correspondence suggests that a \emph{deep} embedding of $\mathsf{G3KGL}$ is unnecessary. Since internalising possible world semantics in sequent calculi is, in fact, a syntactic formalisation of that very semantics, we can use \emph{our own} formalisation in HOL Light of validity in Kripke frames and \emph{adapt} the goal stack mechanism of the theorem prover to develop $\mathsf{G3KGL}$ proofs by relying on that very mechanism.

That adaptation starts with generalising the standard tactics for classical propositional logic already defined in HOL Light.

\medskip

Let us call any expression of forcing in HOL Light notation a \verb|holds|-proposition.

As an abstract deductive system, the logical engine underlying the proof development in HOL Light consists of a single-consequent sequent calculus for higher-order logic. We must work on a multi-consequent sequent calculus made of multisets of \verb|holds|-propositions and (formalised) relational atoms. To handle commas, we recur to the logical operators of HOL Light: a comma in the antecedent is formalised by a conjunction; in the consequent, it is formalised by disjunction.

Since we cannot directly define multisets, we need to formalise the sequent calculus rules to operate on lists and handle permutations through standard conversions of a goal with the general shape of an $n$-ary disjunction of \verb|holds|-propositions, for $n\geq 0$.

The intermediate tactics we now need to define operate \begin{itemize}
\item on the components of a general goal-term consisting of a finite disjunction of \verb|holds|-propositions: these correspond to the labelled formulas that appear on the right-hand side of a sequent of $\mathsf{G3KGL}$ so that some of our tactics can behave as the appropriate right rules of that calculus;
\item on the list of hypotheses of the goal stack, which correspond to the labelled formulas and possible relational atoms that occur on the left-hand side of a sequent of $\mathsf{G3KGL}$, so that some of our tactics can behave as the appropriate left rules of that calculus.
\end{itemize}

In order to make the automation of the process easier, among the hypotheses of each goal stack, we make explicit the assumptions about the transitivity of the accessibility relation, the left-to-right direction of the standard forcing condition for the $\Box$, and, separately, the right-to-left direction of the forcing condition for the same modality in \verb|ITF| models. This way, the formal counterparts of, respectively, $\textit{Trans}$, $\mathcal{L}\Box$, and $\mathcal{R}\Box^{\textit{L\"ob}}$ can be executed adequately by combining standard tactics of HOL Light.

Our classical propositional tactics implement, for purely practical reasons, left and right rules for the (bi) implication-free fragment of $\mathcal{L}$. Therefore, the only intermediate tactics determining a branching in the derivation tree -- i.e.~the generation of subgoals in HOL Light goal-stack -- are those corresponding to $\mathcal{L}\vee$ and $\mathcal{R}\wedge$. The translation of a formula of $\mathcal{L}_{LS}$, given as a goal, to its equivalent formula in the implemented fragment is produced by our conversion \verb+HOLDS_NNFC_UNFOLD_CONV+ \srclinkfoot{swh:1:cnt:4af3d48af0f8b3c2dcef2680c9814f57c342ae2a;origin=https://github.com/jrh13/hol-light;visit=swh:1:snp:fe1b3b83e1dd2bde44d3698f723b25c949de2851;anchor=swh:1:rev:ab57c07ff0105fef75a9fcdd179eda0d26854ba3;path=/GL/decid.ml;lines=128-131}.
A similar conversion is defined for the labelled formulas that appear among the hypotheses, i.e.~on the left-hand side of our formal sequents.

Specific tactics handle left rules for propositional connectives: each is defined by using HOL Light theorem tactic(al)s, which can be thought of as operators on a given goal, taking theorems as input to apply a resulting tactic to the latter. For instance, the left rule for negation $\mathcal{L}\neg$ is defined by
\begin{lstlisting}[mathescape=true,escapeinside={(*!}{!*)}]
let NEG_LEFT_TAC : thm_tactic =(*! \srclink{swh:1:cnt:4af3d48af0f8b3c2dcef2680c9814f57c342ae2a;origin=https://github.com/jrh13/hol-light;visit=swh:1:snp:fe1b3b83e1dd2bde44d3698f723b25c949de2851;anchor=swh:1:rev:ab57c07ff0105fef75a9fcdd179eda0d26854ba3;path=/GL/decid.ml;lines=112-114} !*)
    let pth = MESON [] `$\neg$P $\Longrightarrow$ (P $\vee$ Q) $\Longrightarrow$ Q` in
    MATCH_MP_TAC o MATCH_MP pth
\end{lstlisting}
which uses the propositional tautology $\neg P \Longrightarrow (P \vee Q) \Longrightarrow Q$ in an MP rule instantiated with a negated \verb|holds|-proposition occurring among the hypotheses; then it adds the \verb|holds|-proposition among the disjuncts of the new goal, as expected. $\mathcal{R}\neg$ is defined analogously by \verb|NEG_RIGHT_TAC| \srclinkfoot{swh:1:cnt:4af3d48af0f8b3c2dcef2680c9814f57c342ae2a;origin=https://github.com/jrh13/hol-light;visit=swh:1:snp:fe1b3b83e1dd2bde44d3698f723b25c949de2851;anchor=swh:1:rev:ab57c07ff0105fef75a9fcdd179eda0d26854ba3;path=/GL/decid.ml;lines=101-105}.

On the contrary, $\mathcal{L}\vee$ and $\mathcal{R}\wedge$ are implemented by combining theorem tactic(al)s based on the basic operators \verb|CONJ_TAC| and \verb|DISJ_CASES|.

Modal rules are handled employing the explicit hypotheses we previously mentioned to deal with $\mathcal{L}\Box$ and $\mathcal{R}\Box^{\textit{L\"ob}}$.
The same approach also works for the rule \textit{Trans}, which we handle through a theorem tactical \verb|ACC_TCL| \srclinkfoot{swh:1:cnt:4af3d48af0f8b3c2dcef2680c9814f57c342ae2a;origin=https://github.com/jrh13/hol-light;visit=swh:1:snp:fe1b3b83e1dd2bde44d3698f723b25c949de2851;anchor=swh:1:rev:ab57c07ff0105fef75a9fcdd179eda0d26854ba3;path=/GL/decid.ml;lines=167-170} operating on the relational atoms among the hypotheses of a current goal stack.

This basically completes the formalisation -- or shallow embedding -- of $\mathsf{G3KGL}$ in HOL Light.

\subsection{Design of the proof search}
Our efforts now turn to run, in HOL Light, an automated proof search w.r.t.~the implementation of $\mathsf{G3KGL}$ we have sketched in the previous section.\\
We can again rely on theorem tactic(al)s to build the main algorithm, but we need to define them recursively this time.

First, we have to apply recursively the left rules for propositional connectives, as well as the $\mathcal{L}\Box$ rule: this is made possible by the theorem tactic \verb|HOLDS_TAC| \srclinkfoot{swh:1:cnt:4af3d48af0f8b3c2dcef2680c9814f57c342ae2a;origin=https://github.com/jrh13/hol-light;visit=swh:1:snp:fe1b3b83e1dd2bde44d3698f723b25c949de2851;anchor=swh:1:rev:ab57c07ff0105fef75a9fcdd179eda0d26854ba3;path=/GL/decid.ml;lines=174-176}. Furthermore, we need to saturate the sequents w.r.t.~the latter modal rule, by considering all the possible relational atoms and applications of the rule for transitivity, and, eventually, further left rules: that is the job of the theorem tactic \verb|SATURATE_ACC_TAC| \srclinkfoot{swh:1:cnt:4af3d48af0f8b3c2dcef2680c9814f57c342ae2a;origin=https://github.com/jrh13/hol-light;visit=swh:1:snp:fe1b3b83e1dd2bde44d3698f723b25c949de2851;anchor=swh:1:rev:ab57c07ff0105fef75a9fcdd179eda0d26854ba3;path=/GL/decid.ml;lines=183-185}.

After that, it is possible to proceed with the $\mathcal{R}\Box^\text{Löb}$: that is trigged by the \verb|BOX_RIGHT_TAC| \srclinkfoot{swh:1:cnt:4af3d48af0f8b3c2dcef2680c9814f57c342ae2a;origin=https://github.com/jrh13/hol-light;visit=swh:1:snp:fe1b3b83e1dd2bde44d3698f723b25c949de2851;anchor=swh:1:rev:ab57c07ff0105fef75a9fcdd179eda0d26854ba3;path=/GL/decid.ml;lines=188}, which operates by applying (the implementation of) $\mathcal{R}\Box^\text{Löb}$ after \verb|SATURATE_ACC_TAC| and \verb|HOLDS_TAC|.

At this point, it is possible to optimise the application of \verb|BOX_RIGHT_TAC| by applying the latter tactic after a ``sorting tactic'' \verb|SORT_BOX_TAC| \srclinkfoot{swh:1:cnt:4af3d48af0f8b3c2dcef2680c9814f57c342ae2a;origin=https://github.com/jrh13/hol-light;visit=swh:1:snp:fe1b3b83e1dd2bde44d3698f723b25c949de2851;anchor=swh:1:rev:ab57c07ff0105fef75a9fcdd179eda0d26854ba3;path=/GL/decid.ml;lines=223-227}: that tactic performs a conversion of the goal term and orders it so that priority is given to negated \verb|holds|-propositions, followed by those \verb|holds|-propositions formalising the forcing of a boxed formula. Each of these types of \verb|holds|-propositions are sorted furthermore as follows:

\begin{quote}
\verb|holds WR V p w| precedes \verb|holds WR V q w| if \verb|p| occurs free in \verb|q| and \verb|q| does not occur free in \verb|p|; or if \verb|p| is ``less than'' \verb|q| w.r.t.~the structural ordering of types provided by the OCaml module \verb|Pervasives|.
\end{quote}

We programmed the tactic \verb|GL_TAC| to perform the complete proof search; from that tactic, we define the expected \verb|GL_RULE| \srclinkfoot{swh:1:cnt:4af3d48af0f8b3c2dcef2680c9814f57c342ae2a;origin=https://github.com/jrh13/hol-light;visit=swh:1:snp:fe1b3b83e1dd2bde44d3698f723b25c949de2851;anchor=swh:1:rev:ab57c07ff0105fef75a9fcdd179eda0d26854ba3;path=/GL/decid.ml;lines=270-277}.






To keep it short, our tactic works as expected:
\begin{enumerate}
\item Given a formula $A$ of $\mathcal{L}$, \verb|let|-terms are rewritten together with definable modal operators, and the goal is set to \texttt{|}\verb|-- A|;
\item A model $\langle W,R,V\rangle$ and a world $w\in W$ -- where $W$ sits on the type \verb|num| -- are introduced. The main goal is now \verb|holds (W,R) V A w|; 
\item Explicit additional hypotheses 
are introduced to be able to handle modal and relational rules, as anticipated before;
\item All possible propositional rules are applied after unfolding the modified definition of the predicate \verb|holds| given by \verb|HOLDS_NNFC_UNFOLD_CONV|. This assures that at each step of the proof search, the goal term is a finite conjunction of disjunctions of positive and negative \verb|holds|-propositions. As usual, priority is given to non-branching rules, i.e.\ to those that do not generate subgoals. Furthermore, the hypothesis list is checked, and \textit{Trans} is applied whenever possible; the same holds for $\mathcal{L}\Box$, which is applied any appropriate hypothesis after the tactic triggering transitivity. Each new goal term is reordered by \verb|SORT_BOX_TAC|, which always precedes the implementation of $\mathcal{R}\Box^{\textit{L\"ob}}$.
\end{enumerate}
The procedure is repeated starting from step 2. The tactic governing this repetition is \verb|FIRST o map CHANGED_TAC|, which triggers the correct tactic -- corresponding to a specific step of the very procedure -- in \verb|GL_STEP_TAC| \srclinkfoot{swh:1:cnt:4af3d48af0f8b3c2dcef2680c9814f57c342ae2a;origin=https://github.com/jrh13/hol-light;visit=swh:1:snp:fe1b3b83e1dd2bde44d3698f723b25c949de2851;anchor=swh:1:rev:ab57c07ff0105fef75a9fcdd179eda0d26854ba3;path=/GL/decid.ml;lines=238-241} that does not fail.

At each step, moreover, the following condition is checked by calling \verb|ASM_REWRITE_TAC|:
\begin{description}
\item[\textbf{Closing}] The same \verb|holds|-proposition occurs both among the current hypotheses \emph{and} the disjuncts of the (sub)goal; or \verb|holds (W,R) V False x| occurs in the  current hypothesis list for some label \verb|x|.
\end{description}
This condition states that the current branch is closed, i.e.~an initial sequent has been reached, or the sequent currently analysed has a labelled formula $x:\bot$ in the antecedent.

Termination of the proof search is assured by the results presented in \cite{negri2014proofs}. Therefore, we can justify our choice to conclude \verb|GL_TAC| by a \verb|FAIL_TAC| that, when none of the steps 2--4 can be repeated during a proof search, our algorithm terminates, informing us that a countermodel for the input formula can be built.

That is exactly the job of our \verb|GL_BUILD_COUNTERMODEL| \srclinkfoot{swh:1:cnt:4af3d48af0f8b3c2dcef2680c9814f57c342ae2a;origin=https://github.com/jrh13/hol-light;visit=swh:1:snp:fe1b3b83e1dd2bde44d3698f723b25c949de2851;anchor=swh:1:rev:ab57c07ff0105fef75a9fcdd179eda0d26854ba3;path=/GL/decid.ml;lines=253-264} tactic: it considers the goal state which the previous tactics of \verb|GL_TAC| stopped at; collects all the hypotheses, discarding the meta-hypotheses; and negates all the disjuncts constituting the goal term. Again, by referring to the results in \cite{negri2011proof,negri2014proofs}, we know that this information suffices to the user to construct a relational countermodel for the formula $A$ given as input to our theorem prover for GL.

\subsection{Some examples}

Because of its adequate arithmetical semantics, G\"odel-L\"ob logic reveals an exceptionally simple instrument to study the arithmetical phenomenon of self-reference, as well as G\"odel's results concerning (in)completeness and (un)provability of consistency.

From a formal viewpoint, an arithmetical realisation $\ast$ in Peano arithmetic (\textsf{PA})\footnote{Actually, we may consider any $\Sigma_{1}$-sound arithmetical theory $T$ extending $\mathsf{I\Sigma_{1}}$ \cite{sep-logic-provability}.} of our modal language consists of a function commuting with propositional connectives and such that $(\Box A)^{\ast}:=Bew(\ulcorner A^{\ast}\urcorner)$, where $Bew(x)$ is the formal provability predicate for \textsf{PA}.

Under this interpretation, we will read modal formulas as follows:
\begin{center}

\begin{tabular}{l l}

$\Box A$ & $A$ is provable in \textsf{PA} \\
$\neg \Box \neg A$ & $A$ is consistent with \textsf{PA}                            \\
$\neg\Box A$ & $A$ is unprovable in \textsf{PA}                                     \\
$\Box\neg A$ & $A$ is refutable in \textsf{PA}                                      \\
$(\Box A)\vee (\Box\neg A)$ & $A$ is decidable in \textsf{PA}                         \\
$(\neg\Box A)\wedge(\neg\Box\neg A)$ & $A$ is undecidable in \textsf{PA}             \\
$\Box (A \leftrightarrow B)$ & $A$ and $B$ are equivalent over \textsf{PA}                         \\
$\Box\bot$ & \textsf{PA} is inconsistent              \\
$\neg\Box\bot,\; \Diamond\top$ & \textsf{PA} is consistent\\

\end{tabular}
\end{center}
We tested our procedure \verb|GL_RULE| on some examples.\footnote{The sources of our tests are available in file \verb|GL/tests.ml|.}
First, as a sanity check, we applied it to all the schemas (including axioms but excluding rules) that were initially proven directly in the GL calculus (see our discussion in Section~\ref{sec:GL-lemmas}). In total, 56 such lemmas were (re-)proven by our procedure in about 3.5 seconds.

Next, we used the procedure to prove some other results of meta-mathematical relevance; for instance:
\begin{description}

  \item[Undecidability of consistency.]
    If \textsf{PA} does not prove its inconsistency, then its consistency is undecidable. The corresponding modal formula is
    $$\neg(\Box\Box\bot)\rightarrow\neg(\Box\neg\Box\bot) \wedge\neg(\Box\neg\neg\Box\bot)$$

  \item[Undecidability of G\"odel's formula.]
    The formula stating its own unprovability is undecidable in \textsf{PA}, if the latter does not prove its inconsistency. The corresponding modal formula is
    $$\Box(A \leftrightarrow \neg\Box A) \wedge \neg\Box\Box\bot \rightarrow\neg\Box A \wedge \neg\Box\neg A$$

  \item[Reflection and iterated consistency.]
    $$\Box ((\Box p \rightarrow p) \rightarrow \Diamond \Diamond \top) \rightarrow
      \Diamond \Diamond \top \rightarrow
      \Box p \rightarrow p$$
        \item[Formalised G\"odel's second incompleteness theorem.]
    In \textsf{PA}, the following is provable: If \textsf{PA} is consistent, it cannot prove its own consistency. The corresponding modal formula is
    $$\neg\Box\bot\;\rightarrow\;\neg\Box\Diamond\top$$
For the reader's sake, we describe for this specific example the main steps constituting both the decision procedure (\verb+GL_RULE+) \emph{and} the shallow embedding of $\mathsf{G3KGL}$ in the interactive proof mechanism of HOL Light.

\medskip

First, the goal is set to
\begin{lstlisting}[mathescape=true,escapeinside={(*!}{!*)}]
  `$\text{\textbar}$-- (Not Box False --> Not Box Diam True)`
\end{lstlisting}
Then, by applying the completeness theorem and unfolding and sorting the disjuncts (\verb+HOLDS_NNFC_UNFOLD_CONV+, and \verb+SORT_BOX_TAC+ discussed above), we obtain the following goal stack:
\begin{lstlisting}[mathescape=true,escapeinside={(*!}{!*)}]
    0 [`w IN W`] (w)

  `$\neg$holds (W,R) V (Box Not Box Not True) w $\vee$ 
   holds (W,R) V (Box False) w`
\end{lstlisting}
Next, by using $\mathcal{R}\neg$ (\verb+NEG_RIGHT_TAC+), we have 
\begin{lstlisting}[mathescape=true,escapeinside={(*!}{!*)}]
    0 [`w IN W`] (w)
    1 [`holds (W,R) V (Box Not Box Not True) w`] (holds)

  `holds (W,R) V (Box False) w`
\end{lstlisting}
Now, we can trigger $\mathcal{R}\Box^{\textit{L\"ob}}$ followed by $\mathcal{L}\Box$ (\verb+BOX_RIGHT_TAC+), to obtain:
\begin{lstlisting}[mathescape=true,escapeinside={(*!}{!*)}]
    0 [`w IN W`] (w)
    1 [`holds (W,R) V (Box Not Box Not True) w`] (holds)
    2 [`y IN W`]
    3 [`R w y`] (acc)
    4 [`holds (W,R) V (Box False) y`] (holds)

  `holds (W,R) V (Box Not True) y $\vee$ 
   holds (W,R) V False y`
\end{lstlisting}
By now, the second disjunct is deleted, and by applying $\mathcal{R}\Box^{\textit{L\"ob}}$ followed by $\mathcal{R}\neg$, we are done by $\mathcal{L}\bot$ (closing condition by \verb+ASM_REWRITE_TAC+). 

\medskip

\end{description}

As already discussed, most of these proofs seem to be unfeasible using a generic proof search approach such as the one implemented by \verb|MESON| or \verb|METIS|, either axiomatically or semantically (see Section~\ref{sec:GL-lemmas} and the beginning of Section~\ref{prooftheorygl}, respectively).
Thus, obtaining the formal proof in HOL Light of some of the above four results using more basic techniques can be challenging.
Our procedure can prove them in a few seconds.

\medskip

Finally, we tested the ability of our procedure to find countermodels.
For instance, consider this reflection principle, which is a non-theorem in GL:
$$\Box(\Box p \vee \Box\neg p) \rightarrow (\Box p \vee \Box\neg p)$$
Our procedure fails (meaning it does not produce a theorem) on this formula and warns the user that a countermodel has been constructed.
As expected, the structure that the countermodel constructor \verb+GL_BUILD_COUNTERMODEL+ returns is the one that can be graphically rendered as
$$
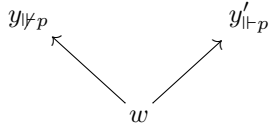
\xymatrix{
y_{\not\Vdash p}  & & y'_{\Vdash p} \\
& w\ar@{->}[ul]\ar@{->}[ur]
}
$$

\section{Related work}\label{relwork}


Our formalisation gives a mechanical proof of completeness for $\mathbb{GL}$ in HOL Light which sticks to the original Henkin's method for classical logic. In its standard version, its nature is synthetic and intrinsically semantic \cite{fittingmendelsohn}. As we stated before, it is the core of the canonical model construction for most normal modal logics.

That approach does not work for $\mathbb{GL}$ because of its non-compactness. This issue is usually sidestepped by restricting to a finite subformula universe, as done in Boolos \cite{boolos1995logic}. Nevertheless, we build the subformula universe in our mechanised proof by recurring to a restricted version of Henkin's construction without resorting to the syntactic manipulations that are proof-sketched in \cite[Ch.~5]{boolos1995logic}.

As far as we know, \emph{no other mechanised proof of modal completeness for} $\mathbb{GL}$ has been given before, despite there exist formalisations of similar results for several other logics.

In particular, both Doczkal and Bard \cite{doczkal2018completeness}, and Doczkal and Smolka \cite{doczkal2016completeness} deal with completeness and decidability of non-compact modal systems -- namely, converse PDL and CTL, respectively. They use Coq/SSReflect to give constructive proofs of completeness on the basis of Kozen and Parikh \cite{KOZEN1981113} for PDL and Emerson and Halpern \cite{EMERSON19851} for CTL. Moreover, they propose a simulation of natural deduction for their formal axiomatic systems to make the proof development of lemmas in the Hilbert calculi that are required in the formalisation of completeness results easier. Their methodology is based on a variant of pruning \cite{kaminski2011correctness}, and the proposed constructive proof of completeness provides an algorithm to build derivations in the appropriate axiomatic calculi for valid formulas.

Formal proof of semantic completeness for \emph{classical logic} has defined an established trend in interactive theorem proving since Shankar \cite{shankar}, where a Hintikka-style strategy is used to define a theoremhood checker for formulas built up by negation and disjunction only.

A very general treatment of systems for classical propositional logic is given in Michaelis and Nipkow \cite{nipkow}. They investigate an axiomatic calculus in Isabelle/HOL along with natural deduction, sequent calculus, and resolution system, and completeness is proven by Hintikka-style method for sequent calculus first, to be lifted then to the other formalisms through translations of each system into the others. Their formalisation is more ambitious than ours, but, at the same time, it is focused on a very different aim. A similar overview of meta-theoretical results for several calculi formalised in Isabelle/HOL is given in Blanchette \cite{blanchette}, who provides a more general investigation, though unrelated to modal logics.

Regarding intuitionistic modalities, Bak \cite{bak} gives a constructive proof of completeness for IS4 w.r.t.~a specific relational semantics verified in Agda. It uses natural deduction and applies modal completeness to obtain a normalisation result for the terms of the associated $\lambda$-calculus.

Bentzen \cite{bentzen} presents a Henkin-style completeness proof for $\mathbb{S}5$ formalised in Lean. That work applies the standard method of canonical models -- since $\mathbb{S}5$ is compact.

More recently, Xu and Norrish \cite{10.1007/978-3-030-51074-9_30} used the HOL4 theorem prover to treat model theory of modal systems.
For future work, it might be interesting to use their formalisation along with the main lines of our implementation of axiomatic calculi to merge the two presentations -- syntactic and semantic -- exhaustively.

\medskip

Our formalisation, however, has been led by the aim of developing a (prototypical) theorem prover in HOL Light for normal modal logics.
The results concerning GL that we have presented here can be considered a case study of our original underlying methodology.

Automated deduction for modal logic has become a relevant scientific activity recently. An exhaustive comparison of our prover with other implementations of modal systems is beyond our scope. Nevertheless, we care to mention at least three different development lines on that trend.

The work \cite{gore2021cegar} by Goré and Kikkert consists of a highly efficient hybridism of SAT solvers, modal clause-learning, and tableaux methods for modal systems. That prover deals with minimal modal logic K and its extensions T and S4. The current version of their implementation does not produce proof or a countermodel for the input formula; however, the code is publicly available, and minor tweaks should make it do so.

The conference paper \cite{girlando2020moin} by Girlando and Stra{\ss}burger presents a theorem prover for intuitionistic modal logics implementing proof search for Tait-style nested sequent calculi in Prolog. Because of the structural properties of those calculi, that prover returns, for each input formula, a proof in the appropriate calculus or a countermodel extracted from the failed proof search in the system. Similar remarks could be formulated for the implementation described in \cite{10.1093/logcom/exab084} concerning several logics for counterfactuals.

The latter formalisations are just two examples of an established modus operandi in implementing proof search in extended sequent calculi for non-classical logics by using the mere depth-ﬁrst search mechanism of Prolog. Other instances of that line are e.g.~\cite{alenda2010csl,giordano2005analytic,giordano2007klmlean}\cite{olivetti2003condlean,olivetti2005condlean,
olivetti2008theorem,olivetti2014nescond}, and \cite{dalmonte:hal-03159954}. None of those provers deals explicitly with GL, but that development approach would find no issue formalising $\mathsf{G3KGL}$ too.


Similar remarks about Papapanagiotou and Fleuriot \cite{papapanagiotou2021object} could be made. They propose a general framework for object-level reasoning with multiset-based sequent calculi in HOL Light. More precisely, they present a \emph{deep} embedding of those systems by defining an appropriate relation between multisets in HOL Light and encode two $\mathsf{G1}$ calculi: a fragment of intuitionistic propositional logic and its Curry-Howard analogous type theory. Specific tactics are then defined to perform an interactive proof search of a given sequent. For our purposes, it might be interesting to check whether their implementation may be of some help to enhance the performance and functionality of our prototypical theorem prover.\footnote{One might say that the framework of \cite{papapanagiotou2021object} is similar to the works in Prolog for aiming a direct deep embedding of a sequent calculus, but it is also close to our implementation for adopting the LCF approach and for choosing HOL Light as the environment.}

It is worth noticing that the paper \cite{10.1007/978-3-030-86059-2_18} by Goré et al.~formalises directly in Coq the $\mathsf{G3}$-style calculus $\mathsf{GLS}$ and proves that the system is structurally complete; most notably, it gives a computer checked proof that the cut rule is admissible in that calculus, and by this formalisation, it clarifies many aspects of the structural analysis of this logic that had remained opaque at the ``pencil and paper'' level.\footnote{We are deeply grateful to an anonymous reviewer for pointing to this work on formalised proof theory for provability logics.}

A specific tableaux based theorem prover for GL is described in Goré and Kelly \cite{gore2007automated}, where they also discuss many efficiency related aspects of their artifact.

 When writing this prototype, we aimed at high flexibility in experimenting with the certification of GL tautologies and with constructing possible countermodels to a modal input formula.
The current stage of the artefact is heavily based on the HOL Light tactic machinery and is slightly naive in some aspects, most notably concerning efficiency. 
A more mature approach would provide a dedicated procedure for the search phase, which uses the tactic mechanism in the last stage of its execution, as done in tactics such as METIS, MESON or ARITH, implemented, for instance, by Harrison \cite{harrison1996optimizing}, Hurd \cite{hurd2003first}, and F{\"{a}}rber and Kaliszyk \cite{farber2015metis}. 

An immediate step would be to enhance the implementation of formal proofs in $\mathsf{G3KGL}$ so that derivation trees are represented as proof objects in the HOL Logic and checked by our procedure. Incidentally, this would constitute an alternative implementation of the labelled sequent calculus using the paradigm of deep embedding instead of shallow embedding. Moreover, we could use it to construct concrete proof trees for $\mathsf{G3KGL}$.
The ideal goal would be to make the prover run in \textsc{PSPACE}.

We also intend to measure the performance of our prototype -- and its eventual definitive version -- using a benchmark set for GL in Balsiger et al. Logics Workbenchs \cite{balsiger2000benchmark}, and compare it with, e.g.~Goré and Kelly's theorem prover.

Moving from the experiments about GL we propose in the present work, we plan to develop a more general mechanism to deal with (ideally) the whole set of normal modal logics within HOL Light. At the same time, we intend to add the necessary machinery to check the generated countermodels without affecting the program's performance.



\section{Declarations}

\subsection{Funding}
Marco Maggesi is supported by the project PRIN2017 “Real and Complex Manifolds: Topology, Geometry and holomorphic dynamics” (code 2017JZ2SW5), and by GNSAGA of INdAM. 

Cosimo Perini Brogi is supported by the project PRIN2017 “IT Matters: Methods and Tools for Trustworthy Smart systems” (code 2017FTXR7S).

\subsection{Authors contribution}
Both authors contributed equally to this work.


\bibliography{GL2}

\end{document}